\documentclass{aa}

\usepackage{graphicx}
\usepackage{txfonts}
\usepackage[colorlinks=true, linkcolor=blue, citecolor=blue, urlcolor=blue]{hyperref}
\usepackage{amsmath}    
\usepackage{float} 
\usepackage{graphicx} 
\usepackage{booktabs} 
\usepackage{hyperref} 
\usepackage{geometry}
\usepackage{ifthen}
\usepackage{lipsum}
\usepackage{color, soul} 
\usepackage{makecell}
\usepackage{subcaption}
\usepackage{xspace} 
\usepackage{multirow}
\usepackage{tablefootnote}
\usepackage{soul}
\usepackage[normalem]{ulem}

\usepackage[normalem]{ulem}
\usepackage[dvipsnames]{xcolor}

\newcommand{\eq}[2][]{\begin{align}
                          #2
\end{align}}

\newcommand{\avg}[1]{\left \langle #1 \right \rangle}
\newcommand{\diff}{\mathrm{d}}

\newcommand{\lr}[1]{\left(#1\right)}

\newcommand{\BH}{\mathrm{BH}}

\newcommand{\nbody}{\texorpdfstring{$N$-body\xspace}{N-body\xspace}}

\newcommand{\units}[1]{\,\mathrm{#1}}

\newcommand{\Gaia}{{\it Gaia}\xspace}

\newcommand{\feh}{\mathrm{[Fe/H]}}
\newcommand{\msun}{{\rm M}_\odot}

\newcommand{\sevn}{\textsc{sevn}\xspace}
\newcommand{\petar}{\textsc{petar}\xspace}
\newcommand{\parsec}{\textsc{parsec}\xspace}
\newcommand{\gaiamock}{\textsc{gaiamock}\xspace}

\def\subinrm#1{\sb{\rm#1}}
    {\catcode`\_=13 \global\let_=\subinrm}
\mathcode`\_="8000
\def\upsubscripts{\catcode`\_=12 } 

\upsubscripts

\newcommand\NSBHZhiDMDRfour{1}
\newcommand\NSBHZloDMDRfour{0.4}
\newcommand\NSBHZhiDMfbDRfour{64}
\newcommand\NSBHZloDMfbDRfour{48}

\newcommand\NSBHZhiDMDRthree{0.2}
\newcommand\NSBHZloDMDRthree{0.04}
\newcommand\NSBHZhiDMfbDRthree{20}
\newcommand\NSBHZloDMfbDRthree{10}

\begin{document}

\title{Dynamical formation of dormant black holes: \\Predictions for \textit{Gaia} DR4}
\titlerunning{Dynamical formation of dormant black holes: Predictions for \textit{Gaia} DR4}

\authorrunning{D. Marín Pina et al.}

\author{Daniel Marín Pina\thanks{e-mail: \href{mailto:daniel.marin@uni-heidelberg.de}{daniel.marin@uni-heidelberg.de}}
\inst{1,2}\orcid{0000-0001-6482-1842}
\and
Michela Mapelli\thanks{e-mail: \href{mailto:mapelli@uni-heidelberg.de}{mapelli@uni-heidelberg.de}}
\inst{1,2,3,4}\orcid{0000-0001-8799-2548}
\and
Giuliano Iorio
\inst{5,6}\orcid{0000-0003-0293-503X}
\and
Cecilia Sgalletta
\inst{1,2}\orcid{0009-0003-7951-4820}
}

\institute{
      Universit\"at Heidelberg, Zentrum f\"ur Astronomie (ZAH), Institut f\"ur Theoretische Astrophysik, Albert-Ueberle-Str. 2, 69120,\\ $^{\,}$ Heidelberg, Germany
    \and
    Universit\"at Heidelberg, Interdisziplin\"ares Zentrum f\"ur Wissenschaftliches Rechnen, Heidelberg, Germany
    \and
    Physics and Astronomy Department Galileo Galilei, University of Padova, Vicolo dell'Osservatorio 3, I--35122, Padova, Italy
    \and
    INAF - Osservatorio Astronomico di Padova, Vicolo dell’Osservatorio 5, I-35122, Padova, Italy
       \and
    Departament de Física, Universitat Politècnica de Catalunya, Av. Eduard Maristany 16, E-08028, Barcelona, Spain
    \and
     Institut de Ciències del Cosmos (ICCUB), Universitat de Barcelona (UB), c. Martí i Franqués, 1, 08028 Barcelona, Spain
}   

\date{Received XX; accepted XX}

\abstract{Dormant black holes (dormant BHs, i.e., X-ray quiet binaries comprising a BH and a companion star) are probes of the effect of cluster dynamics and binary stellar evolution on compact-object binaries. A dozen dormant BH binaries are known, a number that will increase substantially with the upcoming {\it Gaia} Data Release 4 (DR4). Their origin is uncertain, and dynamical formation is proposed as one possible channel. Here, we study the formation of dormant BHs in low-mass star clusters, accounting for the interplay of binary stellar evolution and cluster dynamics. We present new simulations with the $N$-body code {\sc petar}, interfaced with the population synthesis code {\sc sevn} for single and binary star evolution. Our grid spans a range of cluster masses ($300-4800\,$M$_\odot$), metallicities ($1.4\times10^{-2}$, $1.4\times10^{-4}$), and natal kick models (with and without fallback). We find that almost all dormant BH binaries escape the parent cluster within $\sim1\,$Gyr ($\sim10\,$Myr) when assuming weak (strong) natal kicks. Most (73\%) dormant BHs form via the exchange of a BH (or its progenitor star) into a stellar binary, or of a star into a binary with a BH. Dormant BHs in low-mass clusters have a wide orbital period distribution (from a few hours to $10^8\,$d), with an upper limit set by natal kicks, a bimodal eccentricity distribution (a first peak at zero eccentricity for circularised systems and a second at high eccentricity), and stellar masses that follow the (evolved) initial mass function. Large BH masses are a smoking gun of dynamical formation, via exchanges and stellar collisions. We expect {\it Gaia} DR4 to contain $\sim1-64$ high-metallicity and $\sim0.4-48$ low-metallicity dormant BHs, with a period distribution peaking at $\sim10^3\,$d, $\gtrsim90\%$ of them with eccentricity $e>0.1$, and $\sim20\%$ with large BH masses ($m_{\rm BH} \gtrsim 30\,$M$_\odot$).}


\maketitle
%

\section{Introduction}
The known population of black holes (BHs) in binaries reflects the selection biases of the methods used to detect them. X-ray binaries \citep{CorralSantana2015} only reveal systems compact enough ($P\lesssim 10\units{d}$) for the BH to accrete mass from the companion star at a significant rate, enough to be detected in X-rays. A binary BH (BBH) is only visible to current ground-based gravitational-wave (GW) observatories as it merges \citep[][]{TheLIGOScientificCollaboration2025}, leaving wide systems entirely undetected. In both cases, selection effects are significant. Accessing BHs in binaries that are neither X-ray loud nor merging -- `dormant' BHs -- requires different approaches \citep{trimble1969}. One such method is via astrometry and spectroscopy of the companion star in dormant BH binaries. In these binaries, the unseen BH induces a perturbation on the trajectory of the visible star, measurable as anomalies in its parallax, proper motion, and radial velocity.

\textit{Gaia} \citep{Gaia2016, Gaia2018, Gaia2021, Gaia2023} has been instrumental in establishing this population, with three confirmed detections \citep{el-badry_bh2, elbadry_bh1, Chakrabarti2023, Tanikawa2023, GaiaBH3} and multiple other candidates \citep{Andrew2022, Wang2024, MullerHorn2026}.

The available data, as well as the number of additional systems that will be reported in the forthcoming \textit{Gaia} Data Release 4 (DR4), raise the question of the origin of these sources. Current binary-evolution models struggle to reproduce the formation of some of these binaries. For example, the period of \textit{Gaia} BH1 is too short to have formed from the evolution of a non-interacting stellar binary, but also too long to be easily explained as a post-common-envelope system \citep{Rastello2023}. While some binary-evolution models can explain the formation of \textit{Gaia} BH1, they require non-standard assumptions, like high common envelope efficiency \citep{elbadry_bh1}, ad-hoc angular momentum loss only from the donor star during a Roche-lobe overflow \citep{Olejak2026}, or large natal kicks at BH formation \citep{Mapelli2026}.

Of the currently confirmed and candidate dormant BH binaries\footnote{See \href{https://mkenne15.github.io/BHCAT/index.html}{https://mkenne15.github.io/BHCAT} for an updated catalogue.}, five are located in a cluster \citep{Saracino2022, Giesers2018, Giesers2019, Whitaker2026}, one is associated with a stellar stream formed from a dissolved star cluster \citep{GaiaBH3}, and the remaining five are not clearly associated with any stellar structure at the present time \citep{Shenar2022, el-badry_bh2, elbadry_bh1, Mahy2022, Wang2024}, although they might have been in the past \citep[see e.g.,][]{Rastello2023}. One promising avenue to explain the origin of dormant BH binaries is a dynamical formation channel in a dense stellar cluster.

In a cluster, the effect of binary stellar evolution is coupled to dynamical processes. According to the classical dynamical theory of \cite{Heggie1975}, wide (i.e., `soft') binaries tend to become wider after an interaction, while close (i.e., `hard') binaries tend to become closer, with the threshold usually being referred to as the `hard/soft boundary'. This boundary is usually defined as 
\eq{a_{hs}\sim G\frac{m_1m_2}{\epsilon},}
where $a_{hs}$ is the threshold semimajor axis (SMA), $m_{\it i}$ are the component masses, and $\epsilon$ is the mean kinetic energy of a star. In star clusters, hardening of binaries can trigger collisions and mergers, while softening can lead to binary disruption and ionisation. Indeed, dynamical BBH hardening has long been proposed as a pathway for the formation of GW sources \citep[see, e.g.,][]{PortegiesZwart2000, Banerjee2010, Tanikawa2013, Ziosi2014, Mapelli2016, Kremer2020, Kumamoto2020, Banerjee2021b, Chattopadhyay2022, Mapelli2022, Rodriguez2022, FronimosPouliasis2026}. Not only is the orbital period affected, but dynamical interactions can also thermalise the eccentricity distribution, potentially triggering mergers at periapsis, even if the encounter is wide \citep{Heggie1996, Trani2024, RandoForastier2024}.

Strong dynamical encounters can also lead to exchanges in binary components \citep{HillsFullerton1980}. This process preferentially forms binaries with massive components, and so could form a dormant BH from the interaction of a massive BH with a low-mass stellar binary \citep[this has been argued to be the origin of e.g., \textit{Gaia} BH3,][]{MarinPina2026}.

Furthermore, dynamical interactions can lead to the formation of binaries from previously unbound stars and BHs. These are called `three-body' binaries, because their formation requires at least three bodies in a relatively small volume. The properties of these are independent of the initial binary population \citep{Goodman1984, GierszHeggie1994b, Atallah2024, MarinPinaGieles2024, MarinPina2025a}. This has been argued to be the origin of the recently discovered dormant BH in $\omega$ Centauri \citep{Whitaker2026}.

Here, we study the coupled effect of cluster dynamics and single- and binary-stellar evolution processes on the formation of dormant BH binaries. In Sect.~\ref{sec:methods}, we introduce our numerical setup and the initial conditions of our simulation catalogue. In Sect.~\ref{sec:results}, we analyse the dormant BHs found in our simulations, and predict the population of dormant BHs present in \textit{Gaia} DR4. In Sect.~\ref{sec:discussion} and Sect.~\ref{sec:conclusions}, we discuss and summarise our results.

\section{Methods}
\label{sec:methods}
We simulate the coupled effect of cluster dynamics and single- and binary-stellar evolution processes for a population of low-mass young open clusters. We use the \nbody code \petar \citep{Wang2020}, which for the first time we have coupled to the binary population synthesis code \sevn \citep{SperaMapelli2017, Spera2019, Mapelli2020, Iorio2023}.

\subsection{\petar}
\petar \citep{Wang2020} is a high-performance $N$-body code capable of simulating the evolution of dense collisional stellar clusters. It employs a hybrid integration scheme that treats long- and short-range gravitational interactions separately. Long-range interactions are evolved with a Barnes–Hut particle-tree algorithm \citep{BarnesHut1986} combined with a second-order leapfrog integrator, while short-range encounters are handled using a fourth-order Hermite integrator together with slowdown-algorithmic regularisation \citep[][]{SDAR}. This approach enables efficient simulations of star clusters containing a significant binary population.

\subsection{\sevn}
\sevn \citep{SperaMapelli2017, Spera2019, Mapelli2020, Iorio2023} is a population synthesis code that evolves stars and binary stars by interpolating pre-computed stellar tracks. Hence, \sevn allows us to use up-to-date stellar evolution models and to compare different models by changing input evolutionary tracks, without the need to modify the code structure. 
Here, we used the stellar tracks evolved with \parsec, with masses down to $0.7\units{\msun}$ \citep{Bressan2012, Costa2025, Nguyen2025}. Specifically, these tracks include an accurate modelling of stars close to or above a zero-age main sequence mass of 100 M$_\odot$, a large overshooting parameter ($\Lambda_{\rm ov}=0.5$), resulting in larger massive star cores than \textsc{bse} \citep{Hurley2002} models \citep[which is consistent with recent results, e.g.,][]{lechien2026}, and a radiation-driven wind model including the dependence of mass loss on the Eddington ratio $\Gamma=L/L_{\rm Edd}$ (where $L$ and $L_{\rm Edd}$ are the luminosity and Eddington luminosity, respectively). The inclusion of the mass-loss dependence on the Eddington ratio allows to model stellar winds in the optically thick regime -- if a star is radiation-pressure dominated -- and helps massive stars to avoid the Humphreys-Davidson limit \citep{graefener2008,chen2015}. These features result in a different evolution of the stellar radius compared to \textsc{bse} \citep[e.g., Fig.~7 of][]{Iorio2023} and lead to a remarkably different BH mass function even when the same core-collapse supernova model is applied \citep[see][for a detailed discussion and comparison]{Iorio2023}.

We evaluate the properties of stars between $0.1\units{\msun}$ and $0.7\units{\msun}$ using approximate main sequence mass--luminosity and mass--radius scaling relations:
\begin{equation}
     \left(\frac{L}{L_\odot}\right) = \left(\frac{M}{M_\odot}\right)^{3.5}, \qquad
    \left(\frac{R}{R_\odot}\right) = \left(\frac{M}{M_\odot}\right)^{0.8}.
\end{equation}
As the lifetimes of stars in this mass range significantly exceed the age of the Universe, their properties are assumed to be constant during the simulation. Nevertheless, if one of these stars accretes enough matter and overcomes the 0.7M$_\odot$ limit, \sevn interpolates the rest of its life from the \parsec stellar tracks. We refer to \cite{Iorio2023} for an in-depth description of \sevn.

\subsection{Initial conditions}
We construct a catalogue of simulations of low-mass clusters. We vary the initial cluster mass, $M_0$, from $M_0=300\units{\msun}$ to $M_0=4800\units{\msun}$, in steps of a factor of two in mass. Since the formation of dormant BHs is a stochastic process, we generate multiple Monte Carlo realisations of the same initial conditions to filter out stochastic fluctuations. The sampling is performed using \textsc{McLuster} \citep{Kupper2011} for populating the cluster phase space and \textsc{RaccoonBinator} \citep{Torniamenti2021, Mestichelli2026b} to generate the initial binary properties. We ran a larger number of Monte Carlo realisations for lower-mass clusters, ensuring that the total simulated mass for each $M_0$ is the same. This is equivalent to using a cluster initial mass function (CIMF) that scales with the initial cluster mass, $M_0$, as $M_0^{-2}$ \citep{PortegiesZwart2010}. We compute the initial half-mass radius, $r_h$, from \cite{MarksKroupa2012}, i.e., 
\begin{equation}
    \frac{r_h}{1\units{pc}} = 0.1 \lr{\frac{M_0}{1\units{\msun}}}^{0.13},
\end{equation}
which results in initial half-mass densities of $\rho_{h, 0}\sim 10^{3-4}\units{\msun}\units{pc^{-3}}$. We initialise all models with a \cite{King1966} density profile with concentration parameter $W_0=7$, and a \cite{Kroupa2001} initial mass function (IMF) between $0.1\units{\msun}$ and $150\units{\msun}$. The clusters are placed in a Sun-like orbit around the Milky Way (i.e., circular with Galactocentric radius $r_g=8\units{kpc}$), where for the Milky Way potential we used \textsc{MWPotential2014} from \cite{galpy}. 

We initialise our models with a population of binaries constructed from the stars sampled from the IMF, and following the mass-dependent binary fraction of \cite{Moe2017}. Specifically, we follow three different approaches depending on the primary mass. If the primary mass is $m \geq 15\units{\msun}$, we pair the stars up following a \cite{Sana2012} mass ratio distribution, $f(q)\propto q^{-0.1}$, with $0.01\leq q\leq 1$. The periods and eccentricities are sampled from \cite{Sana2012, Moe2017}, with $f(P)\propto (\log P)^{-0.55}$ and $f(e)\propto e^{-0.42}$, with limits set by $10^{0.15}\leq P/(1\units{d})\leq 10^{5.5}$ and $0\leq e \leq 1-[P/(2\units{d})]^{-2/3}$. 

If the primary mass is in the intermediate regime $2<m/(1\units{\msun})<15$, we sample the  orbital parameters from \cite{Kouwenhoven2007}. The mass ratios follow $f(q)\propto q^{-0.4}$, with $0.1\leq q\leq 1$, the eccentricities follow a thermal distribution $f(e)\propto e$, and the period follows $f(P)\propto P^{-1}$ (Öpik's law). 

In the low-mass primary regime, $m\leq 2\units{\msun}$, we sample the initial parameters guided by \cite{Raghavan2010}, with a uniform mass ratio in the range $0.01\leq q\leq 1$ and uniform eccentricity, with periods sampled from a Gaussian distribution in $\log P$, with peak at $\log P/(1\units{d})=5.03$ and standard deviation of $2.28$, truncated to the range $0.1<\log P/(1\units{d})<10$. Overall, the binary fraction is $\simeq 23\%$, and independent of $M_0$.

\subsection{Stellar/binary evolution model parameters}
The usage of \sevn allows us to assume different parametric models for  compact remnant mass, natal kicks, and Roche-lobe efficiency. Here, we describe our assumptions. 

We adopt the rapid supernova (SN) scheme of \cite{Fryer2012} to describe the compact remnant mass function. As for natal kicks, we compare two models: 1) natal kicks randomly sampled from a log-normal distribution with mean value $\mu=5.60$ and standard deviation $\sigma=0.68$ (in km/s), peaking at around $\sim 150 - 200\units{km/s}$, as recently found by \cite{Disberg2025} for Galactic young pulsars (\texttt{*\_DM25}), and 2) natal kicks randomly sampled from the same distribution as above, but then modulated by the fallback fraction (\texttt{*\_DM25fb}), computed as in \cite{Fryer2012}. The choice of the same natal kick distribution for BHs and neutron stars in our \texttt{*\_DM25} models is also motivated by recent studies of core-collapse supernovae supporting large natal kicks for BHs \citep{atri2019,nagarajan2025b,li2026k}.

During mass transfer via Roche-lobe overflow, we assume no accretion onto non-degenerate companions \citep[accretion efficiency $f_a=0$,][]{Hurley2002, Iorio2023}. This process leads to a loss of angular momentum, which we assume is transported from an isotropic wind from the surface of the donor star \citep[so-called Jeans mode, after][]{Jeans1924, Jeans1925}. The assumption of non-conservative mass transfer with Jeans-mode angular-momentum transport is compatible with systems in which the accretor mass is much smaller ($q\leq 0.1$) than the donor mass, and leads to mild orbital shrinking \citep{Olejak2026,Mapelli2026}.

We assume that all stars in the cluster have the same metallicity, and we test two cases: 1) $Z=Z_\odot$ (\texttt{*\_Zhi\_*}, similar to \textit{Gaia} BH1 and BH2) and 2) $Z=0.01 Z_\odot$ (\texttt{*\_Zlo\_*}, similar to \textit{Gaia} BH3). Considering the two natal kick models, the two metallicities, and the five initial star-cluster masses, our catalogue consists of $2\times 2\times 5=20$ unique initial conditions. In addition, we randomly draw multiple realisations of the same initial conditions to filter out stochastic fluctuations. This results in a final catalogue of  1984 cluster simulations. We ran each model for $1\units{Gyr}$, with snapshots taken every $1\units{Myr}$. We summarise the varying catalogue parameters in Table~\ref{table:model_names}.

\begin{table*}[h]
\caption{Summary of the cluster models. From left to right, model name, initial cluster mass $M_0$, initial half-mass radius $r_h$, initial density within the half-mass radius $\rho_{h, 0}$, metallicity $Z$ (where the solar metallicity is $Z_\odot=0.014$), natal kick model, and number of runs performed for each model, $N_{runs}$. The natal kick models DM25 and DM25fb correspond to the distribution by \cite{Disberg2025} applied directly and lowered by the fallback fraction, $f_b$, respectively.}
\label{table:model_names}
\centering 
\begin{tabular}{l c c c c c c}
\hline\hline
Model name & $M_0$ [$\msun$] & $r_h$ [pc] & $\rho_{h, 0}$ [$\msun\units{pc^{-3}}$] & $Z$ & Natal kicks & $N_{runs}$ \\
\hline
\texttt{M03_Zhi_DM25fb} & 300 & 0.210 & $3.87\times 10^{3}$ & 0.014 & DM25fb & 256 \\
\texttt{M03_Zhi_DM25} & 300 & 0.210 & $3.87\times 10^{3}$ & 0.014 & DM25 & 256 \\
\texttt{M03_Zlo_DM25fb} & 300 & 0.210 & $3.87\times 10^{3}$ & 0.00014 & DM25fb & 256 \\
\texttt{M03_Zlo_DM25} & 300 & 0.210 & $3.87\times 10^{3}$ & 0.00014 & DM25 & 256 \\
\texttt{M06_Zhi_DM25fb} & 600 & 0.230 & $5.91\times 10^{3}$ & 0.014 & DM25fb & 128 \\
\texttt{M06_Zhi_DM25} & 600 & 0.230 & $5.91\times 10^{3}$ & 0.014 & DM25 & 128 \\
\texttt{M06_Zlo_DM25fb} & 600 & 0.230 & $5.91\times 10^{3}$ & 0.00014 & DM25fb & 128 \\
\texttt{M06_Zlo_DM25} & 600 & 0.230 & $5.91\times 10^{3}$ & 0.00014 & DM25 & 128 \\
\texttt{M12_Zhi_DM25fb} & 1200 & 0.251 & $9.02\times 10^{3}$ & 0.014 & DM25fb & 64 \\
\texttt{M12_Zhi_DM25} & 1200 & 0.251 & $9.02\times 10^{3}$ & 0.014 & DM25 & 64 \\
\texttt{M12_Zlo_DM25fb} & 1200 & 0.251 & $9.02\times 10^{3}$ & 0.00014 & DM25fb & 64 \\
\texttt{M12_Zlo_DM25} & 1200 & 0.251 & $9.02\times 10^{3}$ & 0.00014 & DM25 & 64 \\
\texttt{M24_Zhi_DM25fb} & 2400 & 0.275 & $1.38\times 10^{4}$ & 0.014 & DM25fb & 32 \\
\texttt{M24_Zhi_DM25} & 2400 & 0.275 & $1.38\times 10^{4}$ & 0.014 & DM25 & 32 \\
\texttt{M24_Zlo_DM25fb} & 2400 & 0.275 & $1.38\times 10^{4}$ & 0.00014 & DM25fb & 32 \\
\texttt{M24_Zlo_DM25} & 2400 & 0.275 & $1.38\times 10^{4}$ & 0.00014 & DM25 & 32 \\
\texttt{M48_Zhi_DM25fb} & 4800 & 0.301 & $2.10\times 10^{4}$ & 0.014 & DM25fb & 16 \\
\texttt{M48_Zhi_DM25} & 4800 & 0.301 & $2.10\times 10^{4}$ & 0.014 & DM25 & 16 \\
\texttt{M48_Zlo_DM25fb} & 4800 & 0.301 & $2.10\times 10^{4}$ & 0.00014 & DM25fb & 16 \\
\texttt{M48_Zlo_DM25} & 4800 & 0.301 & $2.10\times 10^{4}$ & 0.00014 & DM25 & 16 \\
\hline
\end{tabular}
\end{table*}

\subsection{Dormant BH sample selection}
From our simulation catalogue, we extract all dormant BH binaries. \petar does so by applying a KDTree search for nearest neighbours, as in e.g., \cite{Zheng2026}. We select all gravitationally bound pairs containing a non-degenerate star and a BH. From these, we select dormant binaries by requiring that the star does not fill its Roche lobe at periapsis, which we estimate as \citep{Eggleton1983}
\eq{r_{L, \star} = \frac{0.49 q^{2/3}r_p}{0.6q^{2/3}+\ln (1+q^{1/3})},}
with $q=m_\star/m_\BH$ the mass ratio and $r_p$ the periapsis distance. This selection criterion does not exclude wind-fed X-ray binaries (analogues of Cyg X-1). Since \petar does not include GW recoil kicks, we further remove all dormant BHs in which the BH is the product of a BBH merger, as we assume that it has been ejected from the cluster. 

\section{Results}
\label{sec:results}
\subsection{The population of dormant BHs}
\label{sec:population}

Here, we explore the intrinsic population of dormant BH binaries formed in low-mass clusters, without accounting for visibility or selection effects. We average over our cluster snapshots with equal weighting, which would be equivalent to assuming a constant cluster formation rate in the last $1\units{Gyr}$. This choice might under-represent the older systems. 

\subsubsection{Efficiency}
\label{sec:population:eff}

\begin{figure}
    
    \begin{subfigure}{\columnwidth}
    \resizebox{\hsize}{!}{\includegraphics{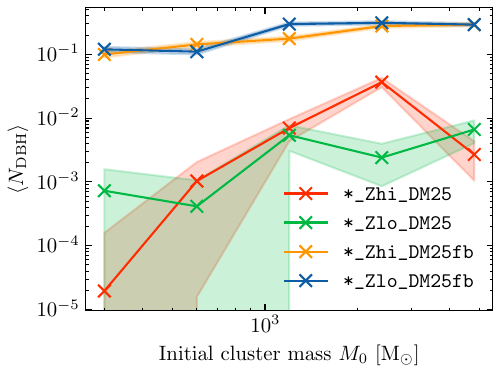}}
    \caption{Average number of dormant BH binaries per cluster, $\avg{N_{DBH}}$}
    \label{fig:NSBH_params}
    \end{subfigure}
    
    \bigskip

    \begin{subfigure}{\columnwidth}
    \resizebox{\hsize}{!}{\includegraphics{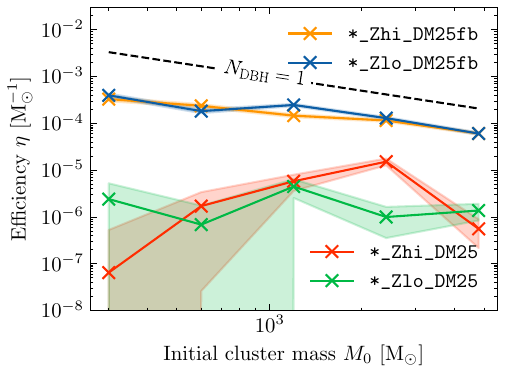}}
    \caption{Dormant BH instantaneous efficiency, $\eta$, defined as the average number of dormant BHs in a cluster at a given timestep divided by the initial star cluster mass ($M_0$). The values of $\eta$ are averaged over snapshots and models.}
    \label{fig:efficiency_params}
    \end{subfigure}

    \caption{ \label{fig:rates} Average number of dormant BHs per cluster (top), and dormant BH instantaneous efficiency (bottom), as a function of the initial cluster mass ($M_0$), separated by metallicity and natal kick model. The shaded region represents the Poissonian uncertainty, estimated as $\sqrt{\sum N_{DBH}}/N_{runs}$.}

\end{figure}

Figure~\ref{fig:NSBH_params} shows the mean number of dormant BHs per cluster, $\avg{N_{DBH}}$, averaged over the different runs, and as a function of the initial cluster mass. We split $\langle N_{DBH}\rangle$ by the initial conditions of the models (masses, metallicities and assumed natal kick prescriptions). In Fig.~\ref{fig:efficiency_params}, we show instead the instantaneous efficiency, defined as
\eq{\eta = \frac{\avg{N_{DBH}}}{M_0},}
where $\avg{N_{DBH}}$ is the average number of dormant BHs in each given snapshot, and $M_0$ is the initial mass of the cluster. We call this `instantaneous' efficiency because it refers to the exact timestep at which we measure it.

Our results are also summarised in Table~\ref{tab:efficiency_params}. The most important parameter determining the number of dormant BH binaries is the assumed natal kick distribution: models in which the kick is lowered by fallback (DM25fb) contain more dormant BHs, because high natal kicks can disrupt the binaries \citep{Fishbach2026}. 

We confirm the finding that $\eta$ has a mild negative correlation with $M_0$, and that on average a cluster hosts $\leq 1$ dormant BH binary \citep{Rastello2023, Tanikawa2024, DiCarlo2024, MarinPina2024}. While lower-mass clusters dominate the production of dormant BHs in the whole Galaxy, a single high-mass cluster has a greater probability of hosting a dormant BH binary than a low-mass cluster. 

The mild negative correlation between the initial cluster mass and $\eta$ can be explained by dynamical interactions leading to binary disruption, which are more relevant at higher cluster masses and densities, as well as dynamical formation of binaries, which is more efficient at lower cluster masses \citep{Goodman1984}. 
In contrast to natal kick and $M_0$, the metallicity of the system does not play a significant role in the instantaneous efficiency.

\begin{table}[h]
\caption{Mean number of dormant BHs, $\avg{N_{DBH}}$, and dormant BH instantaneous efficiency, $\eta$, as a function of the chosen metallicity and natal kick model, averaged over the cluster initial mass function. The errors represent the Poissonian uncertainty.}
\label{tab:efficiency_params}
\centering 
\begin{tabular}{l c c}
\hline\hline
Model name & $\avg{N_{DBH}}$ & $\eta$ [$\msun^ {-1}$] \\
\hline
\texttt{*_Zhi_DM25} & $0.0036 \pm 0.002 $ & $(4.7\pm 3)\times 10^{-6}$ \\
\texttt{*_Zlo_DM25} & $0.0015 \pm 0.001 $ & $(2.0\pm 2)\times 10^{-6}$ \\ 
\texttt{*_Zhi_DM25fb} & $0.14 \pm 0.02 $ & $(1.8 \pm 0.2)\times 10^{-4}$ \\ 
\texttt{*_Zlo_DM25fb} & $0.16 \pm 0.02 $ & $(2.0\pm 0.2)\times 10^{-4}$ \\ 
\hline

\end{tabular}
\end{table}

\begin{figure}
    \resizebox{\hsize}{!}{\includegraphics{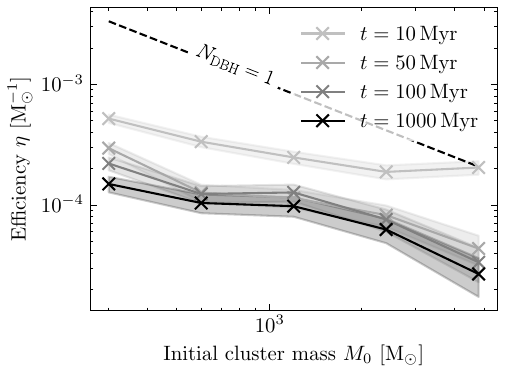}}
    \caption{Dormant BH instantaneous efficiency ($\eta$), as a function of the initial cluster mass ($M_0$), for snapshots taken at specific times after cluster formation. The shaded region represents the Poissonian uncertainty.}
    \label{fig:efficiency_time}
\end{figure}

Finally, we consider the effect of the time evolution of $\eta$. Figure~\ref{fig:efficiency_time} shows $\eta$ as a function of $M_0$, computed from snapshots taken at specific times. After the first $\sim 10\units{Myr}$, we find a much higher instantaneous efficiency, which then drops rapidly. This is because some BBH progenitor binaries (i.e., binaries comprising two stars that will both give birth to a BH) have not yet completed their evolution, and they temporarily appear as dormant BHs. Very young dormant BHs might even have mass ratio $q>1$, i.e., the BH is less massive than the companion.

After the first $\sim 50\units{Myr}$, the instantaneous efficiency becomes almost constant, decreasing only slightly as a few wide binaries are disrupted via dynamical interactions at later times.

\subsubsection{Formation channels}
\label{sec:population:fc}

In this section, we explore the formation channels of dormant BH binaries in a cluster. We first distinguish between primordial and dynamical binaries. Primordial dormant BHs form from the evolution of a stellar binary, possibly with its orbital properties altered by dynamical interactions within the cluster. Dynamical dormant BHs are assembled from an encounter involving a star and a BH that were previously not bound to each other. We can further split the latter into exchanges, if the star and/or BH were previously in a binary; and three-body binaries, if they were not. The interaction that forms a three-body dormant BH needs at least another star and/or BH in close proximity to carry away the excess energy and momentum \citep{Heggie1975, Tanikawa2012}.

We classify a dormant BH as primordial if its components were in a binary in the first snapshot ($t=0$) of the simulation. Otherwise, we take the simulation snapshot before the binary was assembled, and classify it as three-body if its components were single or as exchange if any of them were in a binary.

\begin{figure}
    \resizebox{\hsize}{!}{\includegraphics{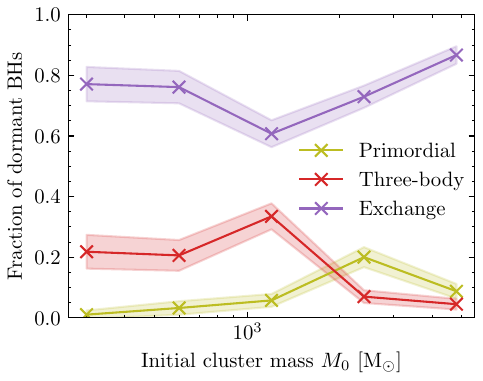}}
    \caption{Fractional contribution of each formation channel (primordial, three-body, and exchange) to the dormant BH population, as a function of the initial cluster mass ($M_0$). Channels are defined as in Sect.~\ref{sec:population:fc}. The shaded region represents the Poissonian uncertainty.}
    \label{fig:form_channel_frac}
\end{figure}

We find that $\sim 6\%$ of our dormant BH binaries are primordial, $\sim 21\%$ are three-body, and $\sim 73\%$ are exchanges. We therefore conclude that exchanges dominate the formation of dormant BHs in clusters. Figure~\ref{fig:form_channel_frac} shows how these fractions depend on the initial cluster mass. Exchanges dominate at all $M_0$, while the three-body contribution drops from $\sim 20\%$ in our lowest-mass clusters to a negligible fraction at higher cluster masses. We can further refine the exchange scenario by checking whether the star and/or the BH (or its progenitor star) were in binaries in the snapshot right before the formation of the dormant BH. In $\sim 37\%$ of the cases the BH (or its progenitor star) was in a binary and exchanged its companion with a single star, in $\sim 43\%$ of the cases the star was in a binary and exchanged its companion with a single BH (or its progenitor star), and in $\sim 20\%$ of the cases both were in binaries and the dormant BH formed via a binary-binary interaction. 

\cite{HillsFullerton1980} showed that, after a strong interaction with null impact parameter involving a binary and a single star/BH, the probability of an exchange is higher if the mass of the intruder is larger than the mass of at least one of the two binary components. As such, we expect that most dormant BHs form from the exchange of a BH into a stellar binary. However, in the simulations we find an almost equal split between a BH exchanging into a binary and a star exchanging into a binary. In the case of BHs exchanging into a binary, according to expectations, we find that they are almost always the most massive of the three bodies ($\sim 99\%$), and most of the time end up pairing with the most massive star ($\sim 62\%$). In the case of stars exchanging into a previously formed dormant BH, the two stars have comparable masses, and the incoming star is the lighter one in $\sim 63\%$ of cases. The previous binary usually has a long period, which increases the cross section, so these exchanges are likely. Then, they tend to harden the resulting dormant BH, with a median shortening of the period of more than an order of magnitude.

We can further characterise the exchanges by separating them depending on whether the already formed BH or its progenitor star exchanged inside a binary. For DM25fb-kick models, $\sim 90\%$ of the dormant BHs form directly with the BH; the interaction does not involve the BH progenitor star. For DM25, we observe the opposite: all dormant BHs form via an exchange involving the BH progenitor star. This is because the natal kick of the BH in the DM25-kick models ejects the dormant BH from the cluster, so it does not undergo further interactions.

As a summary, we find that the formation of dormant BHs in clusters is dominated by exchanges of either a BH (or its progenitor star) into a stellar binary or a star into a previously formed dormant BH.

\subsubsection{Properties of dormant BHs formed in clusters}
\label{sec:population:props}

\begin{figure*}
    \sidecaption
    \includegraphics[width=12cm]{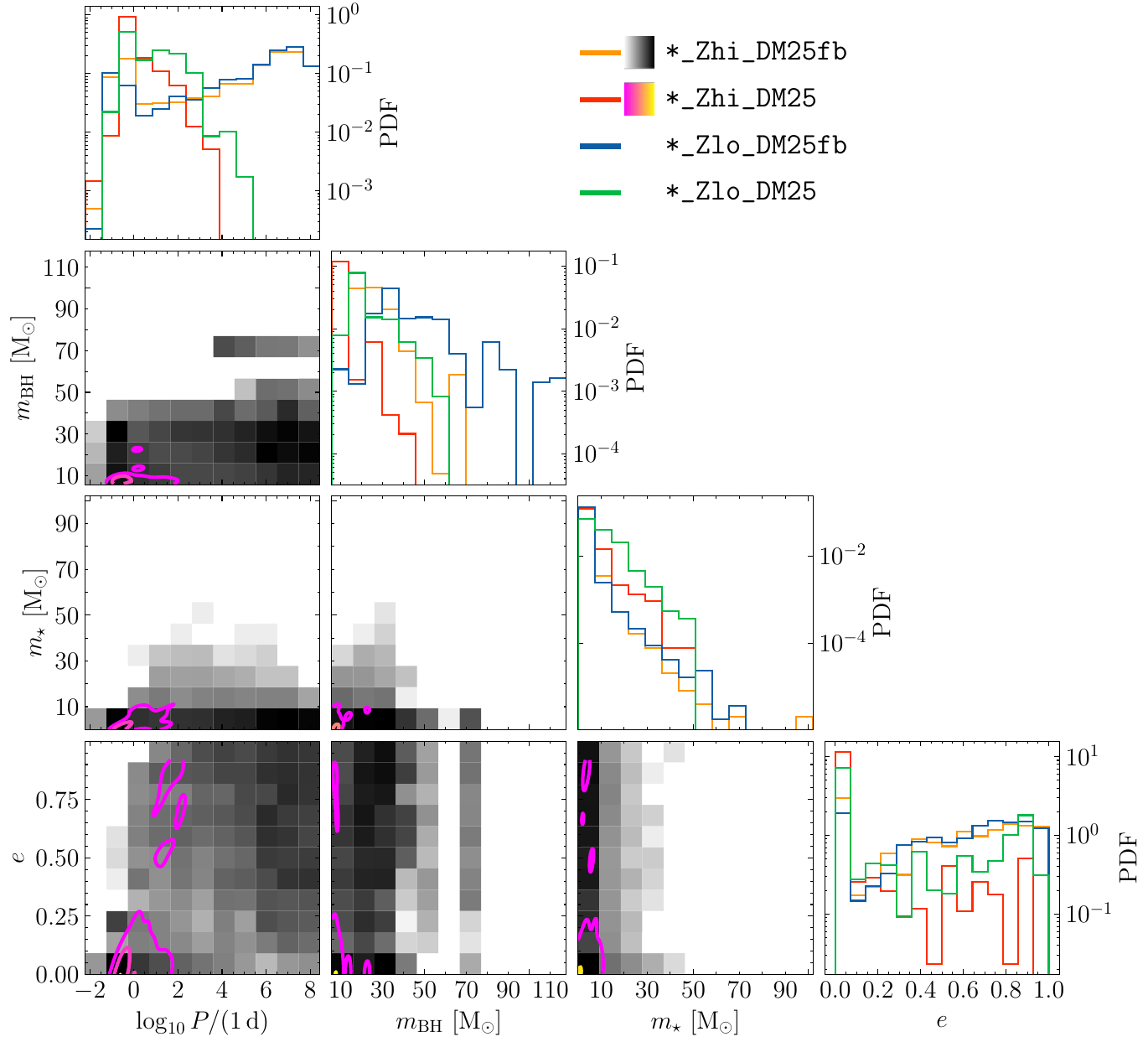}
    \caption{Properties of dormant BHs in our dynamical simulations with high (\texttt{_Zhi}) and low (\texttt{_Zlo}) metallicity and with high (\texttt{_DM25}) and low (\texttt{_DM25fb}) natal kicks. Specifically, we show the probability density function (PDF) of the orbital period ($P$), BH mass ($m_{BH}$), star mass ($m_\star$), and eccentricity ($e$), for all the dormant BHs found in our simulation. The 2D histograms show the high-metallicity models, with DM25fb kicks as a greyscale log-scale gradient, and DM25 kicks as a coloured contour (representing the 20th, 60th, and 90th percentiles). The 1D histograms also include the low-metallicity models.}
    \label{fig:corner:all}
\end{figure*}

Here, we explore the properties of the dormant BHs in our cluster simulations, focusing on the masses, $m_\star$ and $m_\BH$, the orbital period, $P$, and the eccentricity, $e$. We show these parameters, with their correlations, in Fig.~\ref{fig:corner:all}. 

The orbital period distribution is strongly influenced by natal kicks \citep{Mapelli2026}. All models present a peak at $P\sim10^{-0.5}\units{d}$. However, the DM25fb-kick models have a distribution that increases towards larger periods, with a sharp dropoff at $P\gtrsim 10^8\units{d}$, whereas the DM25-kick models (which assume that BHs receive the same natal kicks as neutron stars) show a strong preference for short orbital periods and decline quite fast for $P\ge{}10^3\units{d}$. This can be explained by stronger kicks disrupting wider binaries. 

In our DM25fb-kick models, we find a range of periods comparable to \cite{Rastello2023}, but wider than in \cite{MarinPina2024, DiCarlo2024, Nagarajan2025}. This can be attributed to our models which, like the ones by \cite{Rastello2023}, have lower mass and density than the other works, allowing such long-period binaries to survive long enough in the cluster simulations to be present in our snapshots.

The distribution of the companion mass $m_\star$ scales slightly steeper than $m^{-2.3}_\star$ at large masses, consistent with an evolved IMF. Dormant BHs with massive stars are young; as such, they are dominated by the initial binary population, and most of them are primordial. Dynamical formation channels tend to skew the distribution towards lower $m_\star$, as these are formed later in the cluster evolution, when massive stars have finished their evolution. 

As for the BH mass, we find an anti-correlation of $m_\BH$ to the metallicity. For the DM25-kick models, the kinetic energy provided by the kick scales as $m_\BH$, thus skewing the dormant BH population towards lower $m_\BH$. We also find that the model with low metallicity and DM25fb kicks produces a tail at large BH masses ($\gtrsim 50\units{\msun}$). These BHs are formed after stellar collisions, 
then acquire their companion via dynamical interactions. While in most cases we find $q \leq 1$ (i.e., $m_\BH\geq m_\star$), for $\sim 0.5\%$ of our sample (computed averaging over snapshots) $q>1$, i.e., the star is more massive. These dormant BHs are most likely binary compact object progenitors that have not yet finished their evolution, so if they are found in upcoming surveys, we expect them to be young. We note that the specific fraction of $q>1$ dormant BHs is a decreasing function of the age of the cluster. 

We find a bimodal distribution of orbital eccentricities: a significant fraction of the population ($\sim 17\%$) has near-circular orbits ($e<0.01$); these are mostly post-common-envelope binaries. The rest ($\sim 83\%$) of the population has an eccentricity distribution skewed towards higher $e$; this is a combination of the binaries' initial conditions and the predicted thermal eccentricity distribution for dynamically assembled dormant BHs. As expected, primordial binaries have a much higher fraction of near-circular orbits, with about $69\%$ having $e<0.01$.

\subsubsection{Comparison with isolated binary evolution}
\label{sec:population:iso}

We expect the properties of dormant BH binaries to be affected by dynamical interactions within the cluster, not only if they are dynamically formed, but also if they are primordial binaries. Here, we quantify the difference between the dormant BHs in our simulations and those dormant BHs that we would find if the primordial binaries of the cluster were evolved in isolation. To this purpose, we have extracted all binaries found in our simulations at $t=0$, and evolved them in isolation, using \sevn.  We note that it is not a trivial assumption that binaries formed in clusters share the same initial conditions as those formed in the field, but the exploration of this effect is beyond the scope of this work. We refer to the companion paper by \cite{Mapelli2026} for a study of dormant BH binary formation in isolation.

\begin{figure*}
    \sidecaption
    \includegraphics[width=12cm]{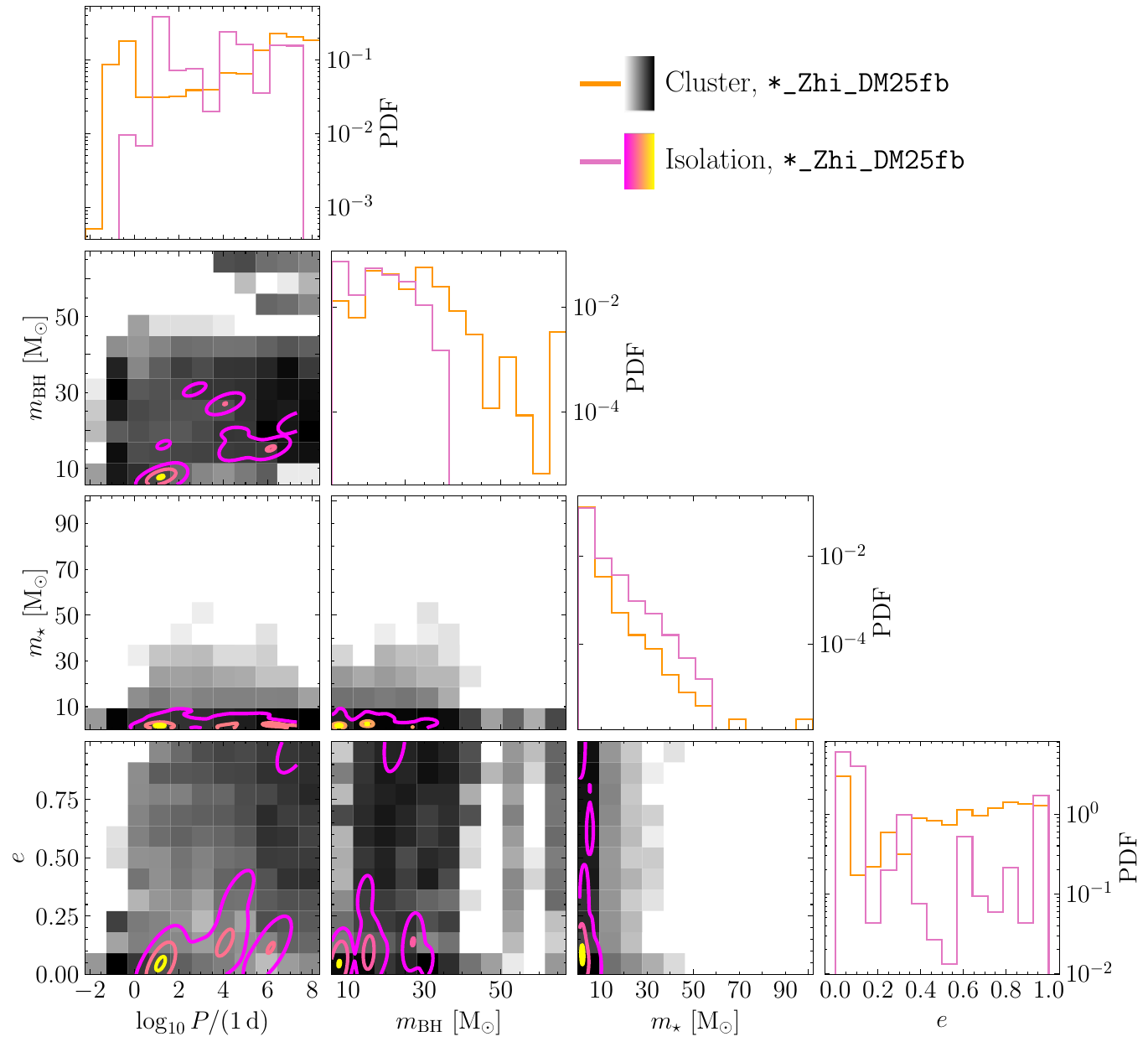}
    \caption{Comparison of dormant BHs in cluster and in isolation. Specifically, we show the PDF of the orbital period ($P$), BH mass ($m_{BH}$), star mass ($m_\star$), and eccentricity ($e$), of dormant BHs formed with high metallicity and DM25fb kicks. In the 2D (1D) histogram, the greyscale log-scale gradient (blue line) shows the dormant BHs in our cluster simulations, whereas the coloured gradient (green line) shows the result of evolving the cluster's primordial binaries in isolation, i.e., unaffected by dynamical interactions. The coloured gradient represents the 20th, 60th, and 90th percentiles.}
    \label{fig:corner:iso}
\end{figure*}

Figure~\ref{fig:corner:iso} compares the dormant BHs evolved in isolation to those found in our cluster simulations, for our model with high metallicity and DM25fb kicks. The properties of dormant BHs evolved in isolation for all the other models are shown in Fig.~\ref{fig:corner:iso_all}. We find that the primordial binaries evolved in isolation show a narrower orbital period distribution. The presence of shorter-period binaries in cluster environments can be explained by dynamical hardening of hard binaries \citep{Heggie1975}. Likewise, wide soft binaries get softer after interactions, leading to the presence of longer-period dormant BHs. Furthermore, dynamical formation (especially after three-body interactions) tends to form wide binaries, which 
are not present in our isolated population. In general, dynamical interactions tend to extend the range of the period distribution to both shorter and longer values.

When comparing the distribution of BH masses, we find that dormant BHs in our cluster simulations have not only a higher median mass than the isolated binaries, but also a high-$m_\BH$ tail not present in our isolated binaries. The most massive BHs in the dynamical simulations are formed from the collision of a stellar binary.  The collision product later becomes a BH and dynamically forms a binary with a companion star. 

The stellar mass distribution reflects the initial conditions -- i.e. the IMF -- in both star clusters and isolation, with a slope that scales slightly steeper than $m_\star^{-2.3}$.

As for the eccentricity, our sample of primordial binaries evolved in isolation has a greater fraction of quasi-circular binaries. This can be explained by the predicted thermal eccentricity distribution ($f(e)=2\,{}e$) after a strong interaction. Furthermore, the presence of dynamical binaries also contributes more to the high-$e$ end of the dormant BH distribution.

\subsubsection{In-cluster and escaped dormant BHs}
\label{sec:population:incl}

Here, we quantify how many dormant BHs remain inside their parent clusters, and explore the differences between in-cluster and escaped dormant BH binaries. We consider a dormant BH to have escaped the cluster if its kinetic energy is greater than the absolute value of the gravitational binding energy to the rest of the cluster, and its distance from the cluster centre is greater than the cluster's tidal radius. 

\begin{figure}
    \resizebox{\hsize}{!}{\includegraphics{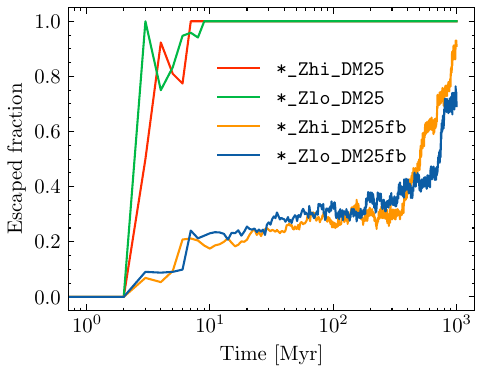}}
    \caption{Fraction of escaped dormant BHs as a function of time, distinguishing metallicity and natal kick models.}
    \label{fig:esc_frac_time}
\end{figure}

Figure~\ref{fig:esc_frac_time} shows the fraction of escaped dormant BHs as a function of time, separated by metallicity and natal kick model. In the DM25 models, almost all dormant BHs are ejected from their parent cluster shortly after their formation because of the large natal kicks. This dramatically quenches their ability to undergo further dynamical interactions. By contrast, only $\sim 20\%$ of the dormant BHs in the DM25fb models are ejected as a consequence of the natal kick. However, also in the DM25fb models, most dormant BHs ($\sim 80-100\%$) become unbound from their parent cluster at $1\units{Gyr}$. Hence, we expect that the vast majority of the dormant BHs dynamically formed in low- and intermediate-mass clusters are now in the Galactic field.

Figure~\ref{fig:corner:esc_incl} shows  the properties of in-cluster and ejected dormant BHs for our model with high metallicity and DM25fb kicks. We find that the properties of these two subpopulations are very similar. The main difference is that in-cluster dormant BHs harbour the most massive BHs, those formed after stellar collisions. These objects are the heaviest components of their parent clusters, so they sink to the centre and the cluster dissolves around them. Another difference is that in-cluster dormant BHs can have more massive star companions; this can be explained by ejected dormant BHs being slightly older (due to the time it takes to eject them). Since our simulations only include low-mass clusters that easily dissolve and have very low escape velocities, we see no significant period cutoff for dynamical ejection. 

We conclude that the population of dormant BHs formed in low-mass clusters and found in streams or in the Galactic field generally matches the (very limited) population of dormant BHs still in their parent clusters.

\subsection{Predictions for \Gaia DR4}
\label{sec:dr4}

Here, we calculate predictions for the rates and properties of observable dormant BHs by applying a realistic forward modelling of the DR4 selection criteria. First, we compute the magnitude of the stars in the \textit{Gaia} $G$ band from the output of our simulations, i.e., the stellar radii and bolometric luminosities. From these, we first compute the bolometric magnitudes, $B$, and effective temperatures, $T_{eff}$; then use the ($T_{eff}$-dependent) $B-V$ colour correction from \cite{Flower1996, Torres2010}, and the $G-V$ correction from \cite{Jordi2010}. 

We then estimate the visibility of each source via the \gaiamock package \citep{gaiamock}. For each dormant BH binary, we sample 100 values of Sun-centric distances in a range $[d_{min}, d_{max}]$, randomly distributed in a disc with an exponential density profile in both Galactocentric radial and perpendicular coordinates (see the \gaiamock documentation for more details). We then assign a visibility value, $f_{v}$, to each dormant BH, which we compute as the fraction of the sampled distances where the dormant BH would be detected in \textit{Gaia} DR4. The absolute value of $f_{v}$ is an ill-defined quantity, as it depends on the minimum and maximum cutoff distance, with $f_{v}\to 0$ as $d_{max}\to\infty$. We circumvent this by repeating this procedure in bins of Sun-centric distance, i.e. $d=[0, 100, 300, 1\,000, 3\,000, 10\,000]\units{pc}$, and computing $f_{v, \textit{i}}$ for each bin. 

\subsubsection{Predicted rates}
\label{sec:dr4:rate}
Using the computed visibility, we now derive an estimate for the number of dormant BHs to be detected in \textit{Gaia} DR4. To do so, we have to take into account the population of open clusters in the local volume. For the masses, we consider a CIMF that scales as $M_0^{-2}$ \citep{PortegiesZwart2010}. Due to our choices in mass resampling, this is equivalent to weighting all simulations equally. Furthermore, we weight the snapshots in our models by open cluster formation times, $t_f$, using the distribution $\phi(t_f)$ reported by \cite{Cavallo2024}. 

Using these distributions, and for each metallicity and natal kick prescription, we compute the average number efficiency $\eta_{N, \textit{i}}$. This quantity represents the number of detected dormant BHs divided by the initial number of stars in the cluster, $N_0$, with the averages taken over CIMF and $\phi(t_f)$, where the visibility is computed for each Sun-centric distance bin $d_{\textit{i}}$. We compute it as follows
\eq{\eta_{N, \textit{i}} = \int \avg{\frac{\sum f_v(t_f, d_{\textit{i}})}{N_0}} \phi(t_f)\,{}\diff t_f,}
where the sum is taken over all dormant BHs present at $t=t_f$ and the average is taken over all models.

In order to obtain a prediction for the number of dormant BHs in DR4, we need to scale this value by the number of stars formed in clusters. 
In DR3 \citep{Gaia2023}, there are $N_{\textit{i}}=[598, 42\,495, 1\,570\,834, 48\,278\,079, 51\,235\,527]$ low-metallicity stars ($-2.5\leq\feh\leq-1.5$) within the distance bins defined in the vector $d$. For high-metallicity stars ($-0.5\leq\feh\leq0.5$), the counts are $N_{\textit{i}}=[94\,375, 3\,012\,836, 26\,923\,299, 46\,162\,252, 28\,104\,779]$. 

\cite{Anders2021} estimated that a fraction $f_c=0.16$ of stars formed in low-mass clusters. Thus, we estimate that DR4 will contain a number of dormant BHs formed in clusters, $N^\mathrm{DR4}_{DBH}$,
\eq{N^\mathrm{DR4}_{DBH}\simeq \sum_{\textit{i}}\,{}\eta_{N, \textit{i}} f_c \,{} N_{\textit{i}}.}
We estimate $N^\mathrm{DR4}_{DBH}\simeq \NSBHZhiDMfbDRfour$ at high metallicity and $N^\mathrm{DR4}_{DBH}\simeq \NSBHZloDMfbDRfour$ at low metallicity for the DM25fb-kick model. For DM25 kicks, we obtain $N^\mathrm{DR4}_{DBH}\simeq \NSBHZhiDMDRfour$ at high metallicity and $N^\mathrm{DR4}_{DBH}\simeq \NSBHZloDMDRfour$ at low metallicity.

To test the validity of our method, we repeat the calculation using the DR3 selection cuts, as implemented in \gaiamock. We estimate $N^\mathrm{DR3}_{DBH}\simeq \NSBHZhiDMfbDRthree$ at high metallicity and $N^\mathrm{DR3}_{DBH}\simeq \NSBHZloDMfbDRthree$ at low metallicity for the DM25fb-kick model. When assuming DM25 kicks, we obtain $N^\mathrm{DR3}_{DBH}\simeq \NSBHZhiDMDRthree$ at high metallicity and $N^\mathrm{DR3}_{DBH}\simeq \NSBHZloDMDRthree$ at low metallicity. For those selection cuts, the actual number of detected dormant BHs is two at high metallicity and none at low metallicity. 

At face value, our DM25fb models predict too many dormant BHs in DR3, especially since this estimate only accounts for the star-cluster formation channel, i.e. does not include isolated field binaries. Moreover, our prediction depends linearly on $f_c$: we recover the observed $N^\mathrm{DR3}_{DBH}$ for models with DM25fb kicks if we assume all dormant BHs are formed in clusters and $f_c\simeq2 \%$ of stars form in low-mass clusters across the Milky Way. Using this value for $f_c$ leads to $N^\mathrm{DR4}_{DBH}\simeq 7$ at high metallicity and $N^\mathrm{DR4}_{DBH}\simeq 5$ at low metallicity for our DM25fb-kick models. Alternatively, the data can also be explained by DM25 kicks, in which case most dormant BHs did not form in star clusters. This might be seen as in tension with the fact that \textit{Gaia} BH3 is associated with a disrupted globular cluster in a tidal stream \citep{Balbinot2024,MarinPina2024, MarinPina2026}. However, our models do not include globular clusters and do not account for accreted dwarf galaxies.

In summary, we predict $N^\mathrm{DR4}_{DBH}\simeq \NSBHZhiDMDRfour-\NSBHZhiDMfbDRfour$ at high metallicity and $N^\mathrm{DR4}_{DBH}\simeq \NSBHZloDMDRfour-\NSBHZloDMfbDRfour$ at low metallicity, with a preference towards the lower values in the range. In Sect.~\ref{sec:discussion:previous}, we compare these predictions to previous work.

\subsubsection{Predicted distribution}
\label{sec:dr4:props}

\begin{figure*}
    \sidecaption
    \includegraphics[width=12cm]{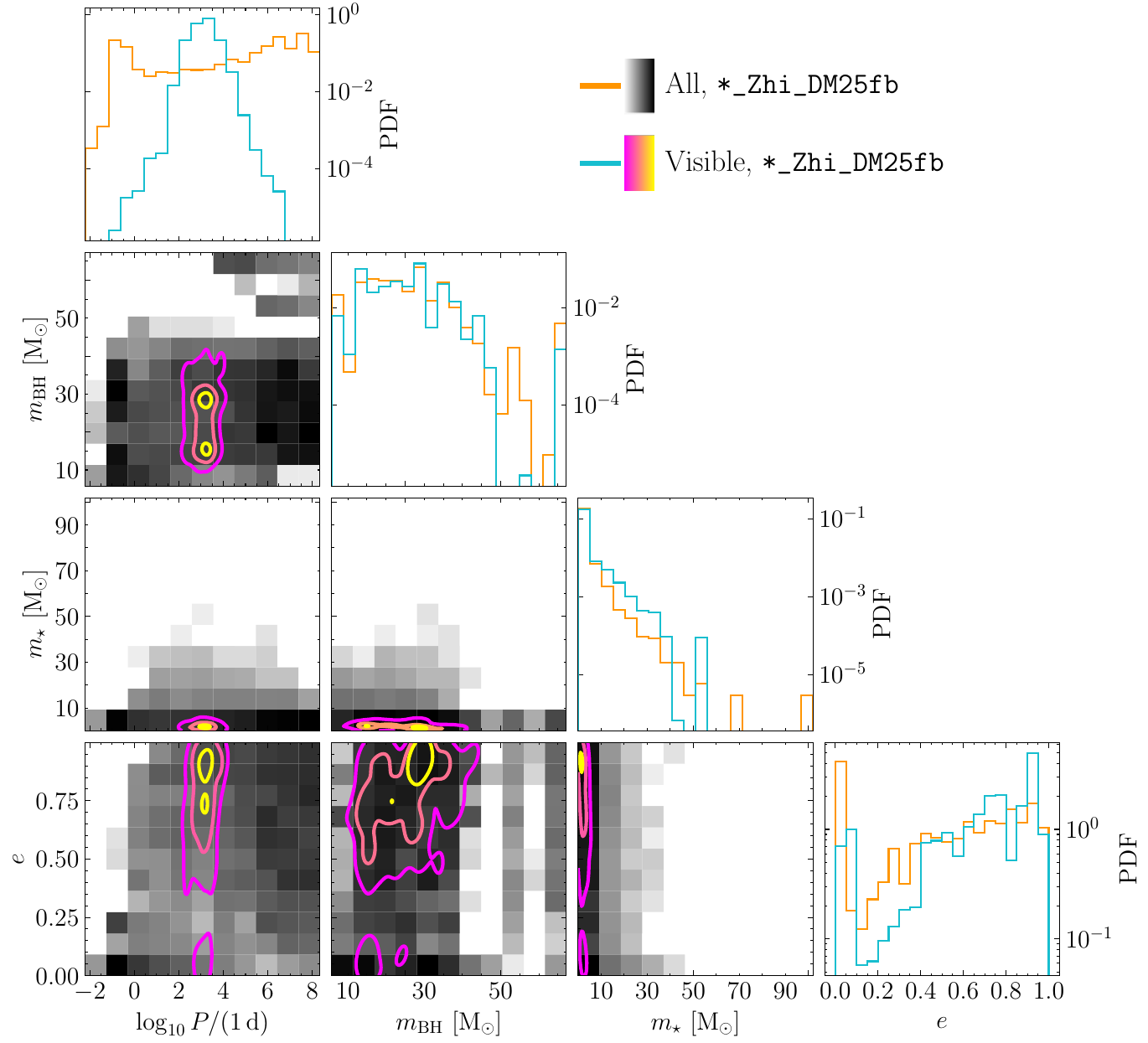}
    \caption{Intrinsic versus detectable properties of dormant BHs. Here, we show the PDF of the orbital period ($P$), BH mass ($m_{BH}$), star mass ($m_\star$), and eccentricity ($e$), for the dormant BHs found in our high metallicity and DM25fb-kick models. In the 2D (1D) histogram, the greyscale log-scale gradient (blue line) shows the inherent dormant BH population, whereas the coloured gradient (green line) shows the population weighted by the detectability in \textit{Gaia} DR4. The coloured gradient represents the 20th, 60th, and 90th percentiles.}
    \label{fig:corner:vis}
\end{figure*}

Here, we estimate the distribution of the properties of dormant BH binaries formed in clusters and detected in \textit{Gaia} DR4. Figure~\ref{fig:corner:vis} shows the distribution of periods, masses, and eccentricities of the intrinsic and visible dormant BH population, for our model with high metallicity and DM25fb kicks. The visibility is estimated by weighting each dormant BH by its $f_{v}$. Figure~\ref{fig:corner:vis_all} shows the visible dormant BH population for all our models, separated by metallicity and natal kick model.

An originally wide distribution in $\log P$ results in a narrowly peaked distribution after detection cuts, with a maximum at $P\sim10^3\units{d}$. As expected, this matches the DR4 observation baseline of $\sim 2\times 10^3\units{d}$. The DM25-kick models have a sharp cutoff at $P\sim 10^5\units{d}$, whereas the DM25fb-kick models extend to longer periods. A comparison of this prediction with the data could be used as supporting evidence for either strong or weak natal kicks, as  discussed by \cite{Mapelli2026} for the isolated binary evolution scenario.

The observed $m_\BH$ distribution closely follows the intrinsic distribution, although the DR4 cuts truncate the higher end of the BH masses. This is not a low-statistic effect, as more massive BHs induce a greater astrometric movement in the star. We predict that $\sim 20\%$ of the detected sample will have $m_\BH\gtrsim 30\units{\msun}$. The DR4 selection cuts tend to mildly favour high-mass stellar companions, due to their higher luminosity. As for orbital eccentricity, we see a preference for detecting eccentric binaries, as their astrometric velocity at periastron is greater than the equivalent circular velocity. We find that $\sim 94\%$ of the detected sample has $e > 0.1$, a higher fraction than in the intrinsic population.

\section{Discussion}
\label{sec:discussion}

Here, we have thoroughly investigated the dynamical formation channel of dormant BHs in star clusters. We find that the possible  `smoking-gun' of dynamical formation is a high BH mass: the most massive BHs in our simulations are stellar collision products that then dynamically form a dormant BH via the interaction with another star(s) and/or binary. Alternatively, a large fraction of eccentric binaries would constitute supporting evidence for a (sub)population of dormant BHs with dynamical origin. The observed period distribution may be too strongly affected by selection effects and uncertainties in binary stellar evolution to fully unearth the difference between dynamical and isolated populations.

While our analysis provides a method to reconstruct the origin of dormant BH populations in DR4, definitive evidence needs to be complemented with independent constraints. For example, the chemical and kinematical association of a dormant BH with a stellar stream provides significant evidence that the dormant BH formed in a cluster environment. Chemical analysis of the surface of the dormant BH star, such as those performed for \textit{Gaia} BH3 \citep{Hackshaw2025, VandenBroeck2026},  can hint towards episodes of mass transfer with the BH progenitor, thus providing support for an isolated formation scenario.

\subsection{Comparison with previous work}
\label{sec:discussion:previous}

Concurrent studies of the formation of dormant BHs in clusters find that these are rare systems, with an expected average number of dormant BH binaries per cluster $\avg{N_{DBH}}\leq 1$ \citep{Rastello2023, MarinPina2024, MarinPina2026, Tanikawa2024, SchiebelbeinZwack2026}. Our predicted period distribution roughly agrees with these studies when assuming DM25fb kicks (none of the previous studies about dormant BHs in star clusters assumes large natal kicks for BHs).

Our predictions for the number of dormant BHs in \textit{Gaia} DR4 (see Sect.~\ref{sec:dr4:rate})  agree with previous work at high metallicity: $\sim 30$ dormant BHs in \cite{Nagarajan2025} (based on the models of \citealt{DiCarlo2024}, but applying a rescaling to fit DR3 data), $\sim 5-130$ in \cite{MarinPina2024} (without taking into account observational cuts). We find significantly more dormant BHs than in the high-mass clusters in \cite{SchiebelbeinZwack2026} (they predict $\sim 0$), which we again explain by the inverse correlation between $\eta$ and $M_0$ (see Sect.~\ref{sec:population}). The models of \cite{DiCarlo2024} may overpredict the number of dormant BHs, which we argue could be due to using lower natal kicks.

\section{Conclusions}
\label{sec:conclusions}
Here, we have presented a model for the population of dormant BH binaries formed in low-mass clusters by the interplay of binary stellar evolution and cluster dynamics. We used the $N$-body code \petar, which we coupled to the binary population synthesis code \sevn, to construct a catalogue of $\sim 2000$ cluster simulations, varying initial cluster masses, metallicities and natal kick prescriptions. \\

\noindent We summarise our main findings as follows.
\begin{itemize}
    \item Dormant BHs are rare systems in clusters, with an expected instantaneous efficiency of $\eta \sim 10^{-6}-10^{-4}\units{\msun^{-1}}$. This value is strongly affected by the assumed natal kick model, with higher kicks leading to lower $\eta$. Lower-mass clusters are more efficient at producing dormant BHs, but a high-mass cluster is more likely to hold a dormant BH than a low-mass cluster.
    \item The formation of dormant BHs in clusters is dominated by exchanges of either a BH (or its progenitor star) into a stellar binary or a star into a previously formed binary with a BH.
    \item We find that the period distribution of dormant BHs has a peak at $P\simeq 10^{-0.5}\units{d}$. In models with natal kicks modulated by fallback (DM25fb), we find a large population of long orbital period systems  (up to $P\simeq 10^8\units{d}$), whereas we find a dropoff at $P \sim 10^{3-4}\units{d}$ for models with strong natal kicks (DM25). This happens because strong natal kicks tend to break long-period binaries. 
    \item The eccentricity is bimodal, with a quasi-circular component and a component skewed towards high eccentricity. 
    \item For some models, we find high BH masses ($\gtrsim 50\units{\msun}$), coming from BHs formed via stellar collisions. The masses of the companion stars follow an evolved IMF. 
    \item Compared to dormant BHs that evolve in isolation (i.e., unaffected by cluster dynamics), we find that dormant BHs formed in clusters have a high-$m_\BH$ tail, a wider period distribution, and a higher fraction of eccentric binaries.
    \item We predict that the vast majority of dormant BHs formed in low- and intermediate-mass clusters escape from their parent cluster either in the first 10 Myr (if natal kicks are strong) or in the first Gyr (if natal kicks are modulated by fallback). However, we find no significant difference between dormant BHs still in-cluster and those that have escaped to the Galactic field, after correcting for the age of the systems. The main difference is that in-cluster systems harbour the most massive BHs in our sample (those formed from stellar collisions), which are not found in our escaped sample. The properties of the vast majority of dormant BHs formed in low-mass clusters do not depend on whether they are ejected or remain inside their parent cluster.
    \item We predict that \textit{Gaia} DR4 will report $\sim \NSBHZhiDMDRfour-\NSBHZhiDMfbDRfour$ high-metallicity dormant BHs and $\sim \NSBHZloDMDRfour-\NSBHZloDMfbDRfour$ low-metallicity dormant BHs formed in low-mass clusters, depending on the assumed natal kick model.
    \item We expect that the population of dormant BHs detected by Gaia DR4 and formed in low-mass clusters will have a period distribution peaked at $P\sim10^3\units{d}$, with $\sim 94\%$ of them having eccentricity $e>0.1$. We also predict that about $\sim 20\%$ of dormant BHs in DR4 will have large BH masses ($m_\BH\gtrsim 30\units{\msun}$).
\end{itemize}

\section*{Data availability}
\sevn is available \href{https://gitlab.com/sevncodes/sevn}{here}.
We have used \sevn v2.16.0 (\texttt{git checkout tags/v2.16.0}). The \petar--\sevn version used in this work is available \href{https://github.com/lwang-astro/PeTar}{here}. The simulations underlying this work will be shared upon reasonable request to the corresponding author.

\begin{acknowledgements}
    The authors thank Gastón Escobar for early advice regarding the \petar-\sevn implementation, and Xiaoying Pang for helpful discussions.
    
    DMP and MM acknowledge support from the Deutsche Forschungsgemeinschaft (DFG, German Research Foundation) through project number 546850815 (acronym: DoBlack, PI: M. Mapelli). DMP, MM  and CS acknowledge support from Germany's Excellence Strategy EXC 2181/1 - 390900948 (the Heidelberg STRUCTURES Excellence Cluster). MM acknowledges support  from the European Research Council for the ERC Advanced grant IMBLACK, under contract no. 101197608. CS acknowledges financial support from the Alexander von Humboldt Foundation for the Humboldt Research Fellowship.
    
    The authors acknowledge support by the state of Baden-Württemberg through bwHPC and the German Research Foundation (DFG) through grants INST 35/1597-1 FUGG 
    %
    and INST 35/1803-1 FUGG and INST 35/1804-1 LAGG.

    This research was supported in part by grant no. NSF PHY-2309135 to the Kavli Institute for Theoretical Physics (KITP). 

    GI acknowledges support from the Ramón y Cajal fellowship RYC2024-048488-I, funded by MICIU/AEI/10.13039/501100011033 and by ESF+

    GI was supported by a fellowship grant from the la Caixa Foundation (ID 100010434). The fellowship code is LCF/BQ/PI24/12040020.
    

\end{acknowledgements}

\bibliographystyle{aa_url} 
\bibliography{bibliography} 

@ARTICLE{chen2015,
       author = {{Chen}, Yang and {Bressan}, Alessandro and {Girardi}, L{\'e}o and {Marigo}, Paola and {Kong}, Xu and {Lanza}, Antonio},
        title = "{PARSEC evolutionary tracks of massive stars up to 350 M$_{☉}$ at metallicities 0.0001 {\ensuremath{\leq}} Z {\ensuremath{\leq}} 0.04}",
      journal = {\mnras},
         year = 2015,
        month = sep,
       volume = {452},
       number = {1},
        pages = {1068-1080},
          doi = {10.1093/mnras/stv1281},
archivePrefix = {arXiv},
       eprint = {1506.01681},
 primaryClass = {astro-ph.SR},
       adsurl = {https://ui.adsabs.harvard.edu/abs/2015MNRAS.452.1068C}
}

@ARTICLE{graefener2008,
       author = {{Gr{\"a}fener}, G. and {Hamann}, W.-R.},
        title = "{Mass loss from late-type WN stars and its Z-dependence. Very massive stars approaching the Eddington limit}",
      journal = {\aap},
         year = 2008,
        month = may,
       volume = {482},
       number = {3},
        pages = {945-960},
          doi = {10.1051/0004-6361:20066176},
archivePrefix = {arXiv},
       eprint = {0803.0866},
 primaryClass = {astro-ph},
       adsurl = {https://ui.adsabs.harvard.edu/abs/2008A&A...482..945G}
}

@ARTICLE{lechien2026,
       author = {{Lechien}, Thibault and {de Mink}, Selma E. and {Justham}, Stephen and {de Burgos}, Abel and {Valli}, Ruggero and {Sim{\'o}n-D{\'\i}az}, Sergio},
        title = "{You're Gonna Need a Bigger Core: Calibrating Massive Star Models against Galactic OB-type Stars}",
      journal = {arXiv e-prints},
         year = 2026,
        month = aug,
          eid = {arXiv:2608.02540},
        pages = {arXiv:2608.02540},
          doi = {10.48550/arXiv.2608.02540},
archivePrefix = {arXiv},
       eprint = {2608.02540},
 primaryClass = {astro-ph.SR},
       adsurl = {https://ui.adsabs.harvard.edu/abs/2026arXiv260802540L}
}

@ARTICLE{Olejak2026,
       author = {{Olejak}, Aleksandra and {Klencki}, Jakub and {Gomez}, Alejandro Vigna- and {de Mink}, Selma E. and {van Son}, Lieke and {Cehula}, Jakub and {Stegmann}, Jakob and {Ryu}, Taeho and {Hendriks}, David D.},
        title = "{Nonconservative Mass Transfer as a Formation Channel for Gaia Black Hole Systems}",
      journal = {\apj},
         year = 2026,
        month = jul,
       volume = {1006},
       number = {1},
          eid = {13},
        pages = {13},
          doi = {10.3847/1538-4357/ae8097},
archivePrefix = {arXiv},
       eprint = {2511.10728},
 primaryClass = {astro-ph.HE},
       adsurl = {https://ui.adsabs.harvard.edu/abs/2026ApJ..1006...13O}
}

@ARTICLE{nagarajan2025b,
       author = {{Nagarajan}, Pranav and {El-Badry}, Kareem},
        title = "{Mixed Origins: Strong Natal Kicks for Some Black Holes and None for Others}",
      journal = {\pasp},
         year = 2025,
        month = mar,
       volume = {137},
       number = {3},
          eid = {034203},
        pages = {034203},
          doi = {10.1088/1538-3873/adb6d6},
archivePrefix = {arXiv},
       eprint = {2411.16847},
 primaryClass = {astro-ph.GA},
       adsurl = {https://ui.adsabs.harvard.edu/abs/2025PASP..137c4203N}
}

@ARTICLE{atri2019,
       author = {{Atri}, P. and {Miller-Jones}, J.~C.~A. and {Bahramian}, A. and {Plotkin}, R.~M. and {Jonker}, P.~G. and {Nelemans}, G. and {Maccarone}, T.~J. and {Sivakoff}, G.~R. and {Deller}, A.~T. and {Chaty}, S. and {Torres}, M.~A.~P. and {Horiuchi}, S. and {McCallum}, J. and {Natusch}, T. and {Phillips}, C.~J. and {Stevens}, J. and {Weston}, S.},
        title = "{Potential kick velocity distribution of black hole X-ray binaries and implications for natal kicks}",
      journal = {\mnras},
         year = 2019,
        month = nov,
       volume = {489},
       number = {3},
        pages = {3116-3134},
          doi = {10.1093/mnras/stz2335},
archivePrefix = {arXiv},
       eprint = {1908.07199},
 primaryClass = {astro-ph.HE},
       adsurl = {https://ui.adsabs.harvard.edu/abs/2019MNRAS.489.3116A}
}

@ARTICLE{li2026k,
       author = {{Li}, Sean E. and {Gottlieb}, Ore and {Metzger}, Brian D.},
        title = "{Strong Black Hole Natal Kicks in Magnetized Accretion-Powered Explosions}",
      journal = {arXiv e-prints},
         year = 2026,
        month = sep,
          eid = {arXiv:2609.10694},
        pages = {arXiv:2609.10694},
          doi = {10.48550/arXiv.2609.10694},
archivePrefix = {arXiv},
       eprint = {2609.10694},
 primaryClass = {astro-ph.HE},
       adsurl = {https://ui.adsabs.harvard.edu/abs/2026arXiv260910694L}
}

@ARTICLE{Trimble1969,
       author = {{Trimble}, Virginia L. and {Thorne}, Kip S.},
        title = "{Spectroscopic Binaries and Collapsed Stars}",
      journal = {\apj},
         year = 1969,
        month = jun,
       volume = {156},
        pages = {1013},
          doi = {10.1086/150032},
       adsurl = {https://ui.adsabs.harvard.edu/abs/1969ApJ...156.1013T}
}

@ARTICLE{HillsFullerton1980,
       author = {{Hills}, J.~G. and {Fullerton}, L.~W.},
        title = "{Computer simulations of close encounters between single stars and hard binaries}",
      journal = {\aj},
         year = 1980,
        month = sep,
       volume = {85},
        pages = {1281-1291},
          doi = {10.1086/112798},
       adsurl = {https://ui.adsabs.harvard.edu/abs/1980AJ.....85.1281H}
}

@ARTICLE{CorralSantana2015,
       author = {{Corral-Santana}, J.~M. and {Casares}, J. and {Mu{\~n}oz-Darias}, T. and {Bauer}, F.~E. and {Mart{\'\i}nez-Pais}, I.~G. and {Russell}, D.~M.},
        title = "{BlackCAT: A catalogue of stellar-mass black holes in X-ray transients}",
      journal = {\aap},
         year = 2016,
        month = mar,
       volume = {587},
          eid = {A61},
        pages = {A61},
          doi = {10.1051/0004-6361/201527130},
archivePrefix = {arXiv},
       eprint = {1510.08869},
 primaryClass = {astro-ph.HE},
       adsurl = {https://ui.adsabs.harvard.edu/abs/2016A&A...587A..61C}
}

@ARTICLE{Eggleton1983,
       author = {{Eggleton}, P.~P.},
        title = "{Aproximations to the radii of Roche lobes.}",
      journal = {\apj},
         year = 1983,
        month = may,
       volume = {268},
        pages = {368-369},
          doi = {10.1086/160960},
       adsurl = {https://ui.adsabs.harvard.edu/abs/1983ApJ...268..368E}
}

@ARTICLE{Disberg2025,
       author = {{Disberg}, Paul and {Mandel}, Ilya},
        title = "{The Kick Velocity Distribution of Isolated Neutron Stars}",
      journal = {\apjl},
         year = 2025,
        month = aug,
       volume = {989},
       number = {1},
          eid = {L8},
        pages = {L8},
          doi = {10.3847/2041-8213/adf286},
archivePrefix = {arXiv},
       eprint = {2505.22102},
 primaryClass = {astro-ph.HE},
       adsurl = {https://ui.adsabs.harvard.edu/abs/2025ApJ...989L...8D}
}

@ARTICLE{Rastello2023,
       author = {{Rastello}, Sara and {Iorio}, Giuliano and {Mapelli}, Michela and {Arca-Sedda}, Manuel and {Di Carlo}, Ugo N. and {Escobar}, Gast{\'o}n J. and {Shenar}, Tomer and {Torniamenti}, Stefano},
        title = "{Dynamical formation of Gaia BH1 in a young star cluster}",
      journal = {\mnras},
         year = 2023,
        month = nov,
       volume = {526},
       number = {1},
        pages = {740-749},
          doi = {10.1093/mnras/stad2757},
archivePrefix = {arXiv},
       eprint = {2306.14679},
 primaryClass = {astro-ph.SR},
       adsurl = {https://ui.adsabs.harvard.edu/abs/2023MNRAS.526..740R}
}

@ARTICLE{Wang2024,
       author = {{Wang}, Song and {Zhao}, Xinlin and {Feng}, Fabo and {Ge}, Hongwei and {Shao}, Yong and {Cui}, Yingzhen and {Gao}, Shijie and {Zhang}, Lifu and {Wang}, Pei and {Li}, Xue and {Bai}, Zhongrui and {Yuan}, Hailong and {Huang}, Yang and {Yuan}, Haibo and {Zhang}, Zhixiang and {Yi}, Tuan and {Xiang}, Maosheng and {Li}, Zhenwei and {Li}, Tanda and {Zhang}, Junbo and {Zhang}, Meng and {Han}, Henggeng and {Fan}, Dongwei and {Li}, Xiangdong and {Chen}, Xuefei and {Liu}, Zhengwei and {Meng}, Xiangcun and {Liu}, Qingzhong and {Zhang}, Haotong and {Gu}, Wei-Min and {Liu}, Jifeng},
        title = "{A potential mass-gap black hole in a wide binary with a circular orbit}",
      journal = {Nature Astronomy},
         year = 2024,
        month = dec,
       volume = {8},
        pages = {1583-1591},
          doi = {10.1038/s41550-024-02359-9},
archivePrefix = {arXiv},
       eprint = {2409.06352},
 primaryClass = {astro-ph.SR},
       adsurl = {https://ui.adsabs.harvard.edu/abs/2024NatAs...8.1583W}
}

@ARTICLE{Whitaker2026,
       author = {{Whitaker}, Matthew and {Kerr}, Evan and {Seth}, Anil and {H{\"a}berle}, Maximilian and {Strader}, Jay and {Anderson}, Jay and {Bellini}, Andrea and {Clontz}, Callie and {Freeman}, Zack and {Griggio}, Massimo and {Kamann}, Sebastian and {Libralato}, Mattia and {Neumayer}, Nadine and {Gonz{\'a}lez Prieto}, Elena and {Rodriguez}, Carl L. and {Saracino}, Sara and {Smith}, Peter and {van de Ven}, Glenn and {Wang}, Zixian},
        title = "{A Long Period Stellar-Mass Black Hole Binary in $ω$ Centauri}",
      journal = {arXiv e-prints},
         year = 2026,
        month = jun,
          eid = {arXiv:2606.18350},
        pages = {arXiv:2606.18350},
archivePrefix = {arXiv},
       eprint = {2606.18350},
 primaryClass = {astro-ph.GA},
       adsurl = {https://ui.adsabs.harvard.edu/abs/2026arXiv260618350W}
}

@ARTICLE{SperaMapelli2017,
       author = {{Spera}, Mario and {Mapelli}, Michela},
        title = "{Very massive stars, pair-instability supernovae and intermediate-mass black holes with the sevn code}",
      journal = {\mnras},
         year = 2017,
        month = oct,
       volume = {470},
       number = {4},
        pages = {4739-4749},
          doi = {10.1093/mnras/stx1576},
archivePrefix = {arXiv},
       eprint = {1706.06109},
 primaryClass = {astro-ph.SR},
       adsurl = {https://ui.adsabs.harvard.edu/abs/2017MNRAS.470.4739S}
}

@ARTICLE{Iorio2023,
       author = {{Iorio}, Giuliano and {Mapelli}, Michela and {Costa}, Guglielmo and {Spera}, Mario and {Escobar}, Gast{\'o}n J. and {Sgalletta}, Cecilia and {Trani}, Alessandro A. and {Korb}, Erika and {Santoliquido}, Filippo and {Dall'Amico}, Marco and {Gaspari}, Nicola and {Bressan}, Alessandro},
        title = "{Compact object mergers: exploring uncertainties from stellar and binary evolution with SEVN}",
      journal = {\mnras},
         year = 2023,
        month = sep,
       volume = {524},
       number = {1},
        pages = {426-470},
          doi = {10.1093/mnras/stad1630},
archivePrefix = {arXiv},
       eprint = {2211.11774},
 primaryClass = {astro-ph.HE},
       adsurl = {https://ui.adsabs.harvard.edu/abs/2023MNRAS.524..426I}
}

@ARTICLE{Mapelli2020,
       author = {{Mapelli}, Michela and {Spera}, Mario and {Montanari}, Enrico and {Limongi}, Marco and {Chieffi}, Alessandro and {Giacobbo}, Nicola and {Bressan}, Alessandro and {Bouffanais}, Yann},
        title = "{Impact of the Rotation and Compactness of Progenitors on the Mass of Black Holes}",
      journal = {\apj},
         year = 2020,
        month = jan,
       volume = {888},
       number = {2},
          eid = {76},
        pages = {76},
          doi = {10.3847/1538-4357/ab584d},
archivePrefix = {arXiv},
       eprint = {1909.01371},
 primaryClass = {astro-ph.HE},
       adsurl = {https://ui.adsabs.harvard.edu/abs/2020ApJ...888...76M}
}

@ARTICLE{Spera2019,
       author = {{Spera}, Mario and {Mapelli}, Michela and {Giacobbo}, Nicola and {Trani}, Alessandro A. and {Bressan}, Alessandro and {Costa}, Guglielmo},
        title = "{Merging black hole binaries with the SEVN code}",
      journal = {\mnras},
         year = 2019,
        month = may,
       volume = {485},
       number = {1},
        pages = {889-907},
          doi = {10.1093/mnras/stz359},
archivePrefix = {arXiv},
       eprint = {1809.04605},
 primaryClass = {astro-ph.HE},
       adsurl = {https://ui.adsabs.harvard.edu/abs/2019MNRAS.485..889S}
}

@ARTICLE{MarinPina2026,
       author = {{Mar{\'\i}n Pina}, Daniel and {Gieles}, Mark and {Rastello}, Sara and {Garcia-Diago}, Cl{\`a}udia and {Iorio}, Giuliano and {Ard{\`e}vol}, Marc},
        title = "{$N$-body modelling of the ED-2 stream progenitor shows Gaia BH3's formation involved dynamical interactions}",
      journal = {arXiv e-prints},
         year = 2026,
        month = apr,
          eid = {arXiv:2604.24874},
        pages = {arXiv:2604.24874},
          doi = {10.48550/arXiv.2604.24874},
archivePrefix = {arXiv},
       eprint = {2604.24874},
 primaryClass = {astro-ph.GA},
       adsurl = {https://ui.adsabs.harvard.edu/abs/2026arXiv260424874M}
}

@ARTICLE{Fishbach2026,
       author = {{Fishbach}, Maya and {Breivik}, Katelyn and {Willcox}, Reinhold and {van Son}, L.~A.~C.},
        title = "{Where are Gaia's small black holes?}",
      journal = {\apss},
         year = 2026,
        month = apr,
       volume = {371},
       number = {4},
          eid = {36},
        pages = {36},
          doi = {10.1007/s10509-026-04554-8},
archivePrefix = {arXiv},
       eprint = {2508.08986},
 primaryClass = {astro-ph.HE},
       adsurl = {https://ui.adsabs.harvard.edu/abs/2026Ap&SS.371...36F}
}

@ARTICLE{Zheng2026,
       author = {{Zheng}, Zepeng and {Wang}, Long and {Baumgardt}, Holger},
        title = "{Binary Disruption during the Early Phase of Open Clusters}",
      journal = {\apj},
         year = 2026,
        month = apr,
       volume = {1001},
       number = {2},
          eid = {225},
        pages = {225},
          doi = {10.3847/1538-4357/ae505a},
archivePrefix = {arXiv},
       eprint = {2603.11803},
 primaryClass = {astro-ph.GA},
       adsurl = {https://ui.adsabs.harvard.edu/abs/2026ApJ..1001..225Z}
}

@ARTICLE{MarksKroupa2012,
       author = {{Marks}, M. and {Kroupa}, P.},
        title = "{Inverse dynamical population synthesis. Constraining the initial conditions of young stellar clusters by studying their binary populations}",
      journal = {\aap},
         year = 2012,
        month = jul,
       volume = {543},
          eid = {A8},
        pages = {A8},
          doi = {10.1051/0004-6361/201118231},
archivePrefix = {arXiv},
       eprint = {1205.1508},
 primaryClass = {astro-ph.GA},
       adsurl = {https://ui.adsabs.harvard.edu/abs/2012A&A...543A...8M}
}

@ARTICLE{PortegiesZwart2010,
       author = {{Portegies Zwart}, Simon F. and {McMillan}, Stephen L.~W. and {Gieles}, Mark},
        title = "{Young Massive Star Clusters}",
      journal = {\araa},
         year = 2010,
        month = sep,
       volume = {48},
        pages = {431-493},
          doi = {10.1146/annurev-astro-081309-130834},
archivePrefix = {arXiv},
       eprint = {1002.1961},
 primaryClass = {astro-ph.GA},
       adsurl = {https://ui.adsabs.harvard.edu/abs/2010ARA&A..48..431P}
}

@ARTICLE{Tanikawa2023,
       author = {{Tanikawa}, Ataru and {Hattori}, Kohei and {Kawanaka}, Norita and {Kinugawa}, Tomoya and {Shikauchi}, Minori and {Tsuna}, Daichi},
        title = "{Search for a Black Hole Binary in Gaia DR3 Astrometric Binary Stars with Spectroscopic Data}",
      journal = {\apj},
         year = 2023,
        month = apr,
       volume = {946},
       number = {2},
          eid = {79},
        pages = {79},
          doi = {10.3847/1538-4357/acbf36},
archivePrefix = {arXiv},
       eprint = {2209.05632},
 primaryClass = {astro-ph.SR},
       adsurl = {https://ui.adsabs.harvard.edu/abs/2023ApJ...946...79T}
}

@ARTICLE{TheLIGOScientificCollaboration2025,
       author = {{The LIGO Scientific Collaboration} and {the Virgo Collaboration} and {the KAGRA Collaboration} and {Abac}, A.~G. and {Abouelfettouh}, I. and {Acernese}, F. and {Ackley}, K. and {Adamcewicz}, C. and {Adhicary}, S. and {Adhikari}, D. and {Adhikari}, N. and {Adhikari}, R.~X. and {Adkins}, V.~K. and {Afroz}, S. and {Agapito}, A. and {Agarwal}, D. and {Agathos}, M. and {Aggarwal}, N. and {Aggarwal}, S. and {Aguiar}, O.~D. and {Ahrend}, I. -L. and {Aiello}, L. and {Ain}, A. and {Ajith}, P. and {Akutsu}, T. and {Albanesi}, S. and {Ali}, W. and {Al-Kershi}, S. and {All{\'e}n{\'e}}, C. and {Allocca}, A. and {Al-Shammari}, S. and {Altin}, P.~A. and {Alvarez-Lopez}, S. and {Amar}, W. and {Amarasinghe}, O. and {Amato}, A. and {Amicucci}, F. and {Amra}, C. and {Ananyeva}, A. and {Anderson}, S.~B. and {Anderson}, W.~G. and {Andia}, M. and {Ando}, M. and {Andr{\'e}s-Carcasona}, M. and {Andri{\'c}}, T. and {Anglin}, J. and {Ansoldi}, S. and {Antelis}, J.~M. and {Antier}, S. and {Aoumi}, M. and {Appavuravther}, E.~Z. and {Appert}, S. and {Apple}, S.~K. and {Arai}, K. and {Araya}, A. and {Araya}, M.~C. and {Arca Sedda}, M. and {Areeda}, J.~S. and {Aritomi}, N. and {Armato}, F. and {Armstrong}, S. and {Arnaud}, N. and {Arogeti}, M. and {Aronson}, S.~M. and {Arun}, K.~G. and {Ashton}, G. and {Aso}, Y. and {Asprea}, L. and {Assiduo}, M. and {Assis de Souza Melo}, S. and {Aston}, S.~M. and {Astone}, P. and {Attadio}, F. and {Aubin}, F. and {AultONeal}, K. and {Avallone}, G. and {Avila}, E.~A. and {Babak}, S. and {Badger}, C. and {Bae}, S. and {Bagnasco}, S. and {Baiotti}, L. and {Bajpai}, R. and {Baka}, T. and {Baker}, A.~M. and {Baker}, K.~A. and {Baker}, T. and {Baldi}, G. and {Baldicchi}, N. and {Ball}, M. and {Ballardin}, G. and {Ballmer}, S.~W. and {Banagiri}, S. and {Banerjee}, B. and {Bankar}, D. and {Baptiste}, T.~M. and {Baral}, P. and {Baratti}, M. and {Barayoga}, J.~C. and {Barish}, B.~C. and {Barker}, D. and {Barman}, N. and {Barneo}, P. and {Barone}, F. and {Barr}, B. and {Barsotti}, L. and {Barsuglia}, M. and {Barta}, D. and {Bartoletti}, A.~M. and {Barton}, M.~A. and {Bartos}, I. and {Basalaev}, A. and {Bassiri}, R. and {Basti}, A. and {Bawaj}, M. and {Baxi}, P. and {Bayley}, J.~C. and {Baylor}, A.~C. and {Baynard}, II, P.~A. and {Bazzan}, M. and {Bedakihale}, V.~M. and {Beirnaert}, F. and {Bejger}, M. and {Belardinelli}, D. and {Bell}, A.~S. and {Bellie}, D.~S. and {Bellizzi}, L. and {Benoit}, W. and {Bentara}, I. and {Bentley}, J.~D. and {Ben Yaala}, M. and {Bera}, S. and {Bergamin}, F. and {Berger}, B.~K. and {Bernuzzi}, S. and {Beroiz}, M. and {Berry}, C.~P.~L. and {Bersanetti}, D. and {Bertheas}, T. and {Bertolini}, A. and {Betzwieser}, J. and {Beveridge}, D. and {Bevilacqua}, G. and {Bevins}, N. and {Bhandare}, R. and {Bhatt}, R. and {Bhattacharjee}, D. and {Bhattacharyya}, S. and {Bhaumik}, S. and {Biancalana}, V. and {Bianchi}, A. and {Bilenko}, I.~A. and {Billingsley}, G. and {Binetti}, A. and {Bini}, S. and {Binu}, C. and {Biot}, S. and {Birnholtz}, O. and {Biscoveanu}, S. and {Bisht}, A. and {Bitossi}, M. and {Bizouard}, M. -A. and {Blaber}, S. and {Blackburn}, J.~K. and {Blagg}, L.~A. and {Blair}, C.~D. and {Blair}, D.~G. and {Bode}, N. and {Boettner}, N. and {Boileau}, G. and {Boldrini}, M. and {Bolingbroke}, G.~N. and {Bolliand}, A. and {Bonavena}, L.~D. and {Bondarescu}, R. and {Bondu}, F. and {Bonilla}, E. and {Bonilla}, M.~S. and {Bonino}, A. and {Bonnand}, R. and {Borchers}, A. and {Borhanian}, S. and {Boschi}, V. and {Bose}, S. and {Bossilkov}, V. and {Bothra}, Y. and {Boudon}, A. and {Bourg}, L. and {Boyle}, M. and {Bozzi}, A. and {Bradaschia}, C. and {Brady}, P.~R. and {Branch}, A. and {Branchesi}, M. and {Braun}, I. and {Briant}, T. and {Brillet}, A. and {Brinkmann}, M. and {Brockill}, P. and {Brockmueller}, E.},
        title = "{GWTC-4.0: Updating the Gravitational-Wave Transient Catalog with Observations from the First Part of the Fourth LIGO-Virgo-KAGRA Observing Run}",
      journal = {arXiv e-prints},
         year = 2025,
        month = aug,
          eid = {arXiv:2508.18082},
        pages = {arXiv:2508.18082},
          doi = {10.48550/arXiv.2508.18082},
archivePrefix = {arXiv},
       eprint = {2508.18082},
 primaryClass = {gr-qc},
       adsurl = {https://ui.adsabs.harvard.edu/abs/2025arXiv250818082T}
}

@ARTICLE{Trani2024,
       author = {{Trani}, Alessandro Alberto and {Leigh}, Nathan W.~C. and {Boekholt}, Tjarda C.~N. and {Portegies Zwart}, Simon},
        title = "{Isles of regularity in a sea of chaos amid the gravitational three-body problem}",
      journal = {arXiv e-prints},
         year = 2024,
        month = mar,
          eid = {arXiv:2403.03247},
        pages = {arXiv:2403.03247},
          doi = {10.48550/arXiv.2403.03247},
archivePrefix = {arXiv},
       eprint = {2403.03247},
 primaryClass = {astro-ph.EP},
       adsurl = {https://ui.adsabs.harvard.edu/abs/2024arXiv240303247T}
}

@ARTICLE{Heggie1975,
       author = {{Heggie}, D.~C.},
        title = "{Binary evolution in stellar dynamics.}",
      journal = {\mnras},
         year = 1975,
        month = dec,
       volume = {173},
        pages = {729-787},
          doi = {10.1093/mnras/173.3.729},
       adsurl = {https://ui.adsabs.harvard.edu/abs/1975MNRAS.173..729H}
}

@ARTICLE{Goodman1984,
       author = {{Goodman}, J.},
        title = "{Homologous evolution of stellar systems after core collapse}",
      journal = {\apj},
         year = 1984,
        month = may,
       volume = {280},
        pages = {298-312},
          doi = {10.1086/161996},
       adsurl = {https://ui.adsabs.harvard.edu/abs/1984ApJ...280..298G}
}

@ARTICLE{King1966,
       author = {{King}, Ivan R.},
        title = "{The structure of star clusters. III. Some simple dynamical models}",
      journal = {\aj},
         year = 1966,
        month = feb,
       volume = {71},
        pages = {64},
          doi = {10.1086/109857},
       adsurl = {https://ui.adsabs.harvard.edu/abs/1966AJ.....71...64K}
}

@ARTICLE{Wang2020,
       author = {{Wang}, Long and {Iwasawa}, Masaki and {Nitadori}, Keigo and {Makino}, Junichiro},
        title = "{PETAR: a high-performance N-body code for modelling massive collisional stellar systems}",
      journal = {\mnras},
         year = 2020,
        month = sep,
       volume = {497},
       number = {1},
        pages = {536-555},
          doi = {10.1093/mnras/staa1915},
archivePrefix = {arXiv},
       eprint = {2006.16560},
 primaryClass = {astro-ph.IM},
       adsurl = {https://ui.adsabs.harvard.edu/abs/2020MNRAS.497..536W}
}

@ARTICLE{Kroupa2001,
       author = {{Kroupa}, Pavel},
        title = "{On the variation of the initial mass function}",
      journal = {\mnras},
         year = 2001,
        month = apr,
       volume = {322},
       number = {2},
        pages = {231-246},
          doi = {10.1046/j.1365-8711.2001.04022.x},
archivePrefix = {arXiv},
       eprint = {astro-ph/0009005},
 primaryClass = {astro-ph},
       adsurl = {https://ui.adsabs.harvard.edu/abs/2001MNRAS.322..231K}
}

@ARTICLE{Kupper2011,
       author = {{K{\"u}pper}, Andreas H.~W. and {Maschberger}, Thomas and {Kroupa}, Pavel and {Baumgardt}, Holger},
        title = "{Mass segregation and fractal substructure in young massive clusters - I. The McLuster code and method calibration}",
      journal = {\mnras},
         year = 2011,
        month = nov,
       volume = {417},
       number = {3},
        pages = {2300-2317},
          doi = {10.1111/j.1365-2966.2011.19412.x},
archivePrefix = {arXiv},
       eprint = {1107.2395},
 primaryClass = {astro-ph.GA},
       adsurl = {https://ui.adsabs.harvard.edu/abs/2011MNRAS.417.2300K}
}

@ARTICLE{PortegiesZwart2000,
       author = {{Portegies Zwart}, Simon F. and {McMillan}, Stephen L.~W.},
        title = "{Black Hole Mergers in the Universe}",
      journal = {\apjl},
         year = 2000,
        month = jan,
       volume = {528},
       number = {1},
        pages = {L17-L20},
          doi = {10.1086/312422},
archivePrefix = {arXiv},
       eprint = {astro-ph/9910061},
 primaryClass = {astro-ph},
       adsurl = {https://ui.adsabs.harvard.edu/abs/2000ApJ...528L..17P}
}

@ARTICLE{Rodriguez2022,
       author = {{Rodriguez}, Carl L. and {Weatherford}, Newlin C. and {Coughlin}, Scott C. and {Amaro-Seoane}, Pau and {Breivik}, Katelyn and {Chatterjee}, Sourav and {Fragione}, Giacomo and {K{\i}ro{\u{g}}lu}, Fulya and {Kremer}, Kyle and {Rui}, Nicholas Z. and {Ye}, Claire S. and {Zevin}, Michael and {Rasio}, Frederic A.},
        title = "{Modeling Dense Star Clusters in the Milky Way and beyond with the Cluster Monte Carlo Code}",
      journal = {\apjs},
         year = 2022,
        month = feb,
       volume = {258},
       number = {2},
          eid = {22},
        pages = {22},
          doi = {10.3847/1538-4365/ac2edf},
archivePrefix = {arXiv},
       eprint = {2106.02643},
 primaryClass = {astro-ph.GA},
       adsurl = {https://ui.adsabs.harvard.edu/abs/2022ApJS..258...22R}
}

@ARTICLE{Hurley2002,
       author = {{Hurley}, Jarrod R. and {Tout}, Christopher A. and {Pols}, Onno R.},
        title = "{Evolution of binary stars and the effect of tides on binary populations}",
      journal = {\mnras},
         year = 2002,
        month = feb,
       volume = {329},
       number = {4},
        pages = {897-928},
          doi = {10.1046/j.1365-8711.2002.05038.x},
archivePrefix = {arXiv},
       eprint = {astro-ph/0201220},
 primaryClass = {astro-ph},
       adsurl = {https://ui.adsabs.harvard.edu/abs/2002MNRAS.329..897H}
}

@ARTICLE{Chattopadhyay2022,
       author = {{Chattopadhyay}, Debatri and {Hurley}, Jarrod and {Stevenson}, Simon and {Raidani}, Arihant},
        title = "{Dynamical double black holes and their host cluster properties}",
      journal = {\mnras},
         year = 2022,
        month = jul,
       volume = {513},
       number = {3},
        pages = {4527-4555},
          doi = {10.1093/mnras/stac1163},
archivePrefix = {arXiv},
       eprint = {2202.08924},
 primaryClass = {astro-ph.GA},
       adsurl = {https://ui.adsabs.harvard.edu/abs/2022MNRAS.513.4527C}
}

@ARTICLE{GierszHeggie1994b,
       author = {{Giersz}, M. and {Heggie}, D.~C.},
        title = "{Statistics of N-Body Simulations - Part Two - Equal Masses after Core Collapse}",
      journal = {\mnras},
         year = 1994,
        month = sep,
       volume = {270},
        pages = {298},
          doi = {10.1093/mnras/270.2.298},
archivePrefix = {arXiv},
       eprint = {astro-ph/9403024},
 primaryClass = {astro-ph},
       adsurl = {https://ui.adsabs.harvard.edu/abs/1994MNRAS.270..298G}
}

@ARTICLE{Moe2017,
       author = {{Moe}, Maxwell and {Di Stefano}, Rosanne},
        title = "{Mind Your Ps and Qs: The Interrelation between Period (P) and Mass-ratio (Q) Distributions of Binary Stars}",
      journal = {\apjs},
         year = 2017,
        month = jun,
       volume = {230},
       number = {2},
          eid = {15},
        pages = {15},
          doi = {10.3847/1538-4365/aa6fb6},
archivePrefix = {arXiv},
       eprint = {1606.05347},
 primaryClass = {astro-ph.SR},
       adsurl = {https://ui.adsabs.harvard.edu/abs/2017ApJS..230...15M}
}

@ARTICLE{Kremer2020,
       author = {{Kremer}, Kyle and {Spera}, Mario and {Becker}, Devin and {Chatterjee}, Sourav and {Di Carlo}, Ugo N. and {Fragione}, Giacomo and {Rodriguez}, Carl L. and {Ye}, Claire S. and {Rasio}, Frederic A.},
        title = "{Populating the Upper Black Hole Mass Gap through Stellar Collisions in Young Star Clusters}",
      journal = {\apj},
         year = 2020,
        month = nov,
       volume = {903},
       number = {1},
          eid = {45},
        pages = {45},
          doi = {10.3847/1538-4357/abb945},
archivePrefix = {arXiv},
       eprint = {2006.10771},
 primaryClass = {astro-ph.HE},
       adsurl = {https://ui.adsabs.harvard.edu/abs/2020ApJ...903...45K}
}

@ARTICLE{BarnesHut1986,
       author = {{Barnes}, Josh and {Hut}, Piet},
        title = "{A hierarchical O(N log N) force-calculation algorithm}",
      journal = {\nat},
         year = 1986,
        month = dec,
       volume = {324},
       number = {6096},
        pages = {446-449},
          doi = {10.1038/324446a0},
       adsurl = {https://ui.adsabs.harvard.edu/abs/1986Natur.324..446B}
}

@ARTICLE{Banerjee2021b,
       author = {{Banerjee}, Sambaran},
        title = "{Stellar-mass black holes in young massive and open stellar clusters - V. comparisons with LIGO-Virgo merger rate densities}",
      journal = {\mnras},
         year = 2021,
        month = may,
       volume = {503},
       number = {3},
        pages = {3371-3385},
          doi = {10.1093/mnras/stab591},
archivePrefix = {arXiv},
       eprint = {2011.07000},
 primaryClass = {astro-ph.HE},
       adsurl = {https://ui.adsabs.harvard.edu/abs/2021MNRAS.503.3371B}
}

@ARTICLE{Tanikawa2012,
       author = {{Tanikawa}, Ataru and {Hut}, Piet and {Makino}, Junichiro},
        title = "{Unexpected formation modes of the first hard binary in core collapse}",
      journal = {\na},
         year = 2012,
        month = apr,
       volume = {17},
       number = {3},
        pages = {272-280},
          doi = {10.1016/j.newast.2011.09.001},
archivePrefix = {arXiv},
       eprint = {1107.3866},
 primaryClass = {astro-ph.GA},
       adsurl = {https://ui.adsabs.harvard.edu/abs/2012NewA...17..272T}
}

@ARTICLE{Kumamoto2020,
       author = {{Kumamoto}, Jun and {Fujii}, Michiko S. and {Tanikawa}, Ataru},
        title = "{Merger rate density of binary black holes formed in open clusters}",
      journal = {\mnras},
         year = 2020,
        month = jul,
       volume = {495},
       number = {4},
        pages = {4268-4278},
          doi = {10.1093/mnras/staa1440},
archivePrefix = {arXiv},
       eprint = {2001.10690},
 primaryClass = {astro-ph.HE},
       adsurl = {https://ui.adsabs.harvard.edu/abs/2020MNRAS.495.4268K}
}

@ARTICLE{Mapelli2022,
       author = {{Mapelli}, Michela and {Bouffanais}, Yann and {Santoliquido}, Filippo and {Arca Sedda}, Manuel and {Artale}, M. Celeste},
        title = "{The cosmic evolution of binary black holes in young, globular, and nuclear star clusters: rates, masses, spins, and mixing fractions}",
      journal = {\mnras},
         year = 2022,
        month = apr,
       volume = {511},
       number = {4},
        pages = {5797-5816},
          doi = {10.1093/mnras/stac422},
archivePrefix = {arXiv},
       eprint = {2109.06222},
 primaryClass = {astro-ph.HE},
       adsurl = {https://ui.adsabs.harvard.edu/abs/2022MNRAS.511.5797M}
}

@ARTICLE{SDAR,
       author = {{Wang}, Long and {Nitadori}, Keigo and {Makino}, Junichiro},
        title = "{A slow-down time-transformed symplectic integrator for solving the few-body problem}",
      journal = {\mnras},
         year = 2020,
        month = apr,
       volume = {493},
       number = {3},
        pages = {3398-3411},
          doi = {10.1093/mnras/staa480},
archivePrefix = {arXiv},
       eprint = {2002.07938},
 primaryClass = {astro-ph.EP},
       adsurl = {https://ui.adsabs.harvard.edu/abs/2020MNRAS.493.3398W}
}

@ARTICLE{Jordi2010,
       author = {{Jordi}, C. and {Gebran}, M. and {Carrasco}, J.~M. and {de Bruijne}, J. and {Voss}, H. and {Fabricius}, C. and {Knude}, J. and {Vallenari}, A. and {Kohley}, R. and {Mora}, A.},
        title = "{Gaia broad band photometry}",
      journal = {\aap},
         year = 2010,
        month = nov,
       volume = {523},
          eid = {A48},
        pages = {A48},
          doi = {10.1051/0004-6361/201015441},
archivePrefix = {arXiv},
       eprint = {1008.0815},
 primaryClass = {astro-ph.IM},
       adsurl = {https://ui.adsabs.harvard.edu/abs/2010A&A...523A..48J}
}

@ARTICLE{MarinPinaGieles2024,
       author = {{Mar{\'\i}n Pina}, Daniel and {Gieles}, Mark},
        title = "{Demographics of three-body binary black holes in star clusters: implications for gravitational waves}",
      journal = {\mnras},
         year = 2024,
        month = jan,
       volume = {527},
       number = {3},
        pages = {8369-8381},
          doi = {10.1093/mnras/stad3777},
archivePrefix = {arXiv},
       eprint = {2308.10318},
 primaryClass = {astro-ph.GA},
       adsurl = {https://ui.adsabs.harvard.edu/abs/2024MNRAS.527.8369M}
}

@ARTICLE{elbadry_bh1,
       author = {{El-Badry}, Kareem and {Rix}, Hans-Walter and {Quataert}, Eliot and {Howard}, Andrew W. and {Isaacson}, Howard and {Fuller}, Jim and {Hawkins}, Keith and {Breivik}, Katelyn and {Wong}, Kaze W.~K. and {Rodriguez}, Antonio C. and {Conroy}, Charlie and {Shahaf}, Sahar and {Mazeh}, Tsevi and {Arenou}, Fr{\'e}d{\'e}ric and {Burdge}, Kevin B. and {Bashi}, Dolev and {Faigler}, Simchon and {Weisz}, Daniel R. and {Seeburger}, Rhys and {Almada Monter}, Silvia and {Wojno}, Jennifer},
        title = "{A Sun-like star orbiting a black hole}",
      journal = {\mnras},
         year = 2023,
        month = jan,
       volume = {518},
       number = {1},
        pages = {1057-1085},
          doi = {10.1093/mnras/stac3140},
archivePrefix = {arXiv},
       eprint = {2209.06833},
 primaryClass = {astro-ph.SR},
       adsurl = {https://ui.adsabs.harvard.edu/abs/2023MNRAS.518.1057E}
}

@ARTICLE{Tanikawa2024,
       author = {{Tanikawa}, Ataru and {Cary}, Savannah and {Shikauchi}, Minori and {Wang}, Long and {Fujii}, Michiko S.},
        title = "{Compact binary formation in open star clusters - I. High formation efficiency of Gaia BHs and their multiplicities}",
      journal = {\mnras},
         year = 2024,
        month = jan,
       volume = {527},
       number = {2},
        pages = {4031-4039},
          doi = {10.1093/mnras/stad3294},
archivePrefix = {arXiv},
       eprint = {2303.05743},
 primaryClass = {astro-ph.GA},
       adsurl = {https://ui.adsabs.harvard.edu/abs/2024MNRAS.527.4031T}
}

@ARTICLE{DiCarlo2024,
       author = {{Di Carlo}, Ugo Niccol{\`o} and {Agrawal}, Poojan and {Rodriguez}, Carl L. and {Breivik}, Katelyn},
        title = "{Young Star Clusters Dominate the Production of Detached Black Hole{\textendash}Star Binaries}",
      journal = {\apj},
         year = 2024,
        month = apr,
       volume = {965},
       number = {1},
          eid = {22},
        pages = {22},
          doi = {10.3847/1538-4357/ad2f2c},
archivePrefix = {arXiv},
       eprint = {2306.13121},
 primaryClass = {astro-ph.GA},
       adsurl = {https://ui.adsabs.harvard.edu/abs/2024ApJ...965...22D}
}

@ARTICLE{GaiaBH3,
       author = {{Gaia Collaboration} and {Panuzzo}, P. and {Mazeh}, T. and {Arenou}, F. and {Holl}, B. and {Caffau}, E. and {Jorissen}, A. and {Babusiaux}, C. and {Gavras}, P. and {Sahlmann}, J. and {Bastian}, U. and {Wyrzykowski}, {\L}. and {Eyer}, L. and {Leclerc}, N. and {Bauchet}, N. and {Bombrun}, A. and {Mowlavi}, N. and {Seabroke}, G.~M. and {{\v{Z}}upi{\'c}}, A. and {Zwitter}, T.},
        title = "{Discovery of a dormant 33 solar-mass black hole in pre-release Gaia astrometry}",
      journal = {\aap},
         year = 2024,
        month = jun,
       volume = {686},
          eid = {L2},
        pages = {L2},
          doi = {10.1051/0004-6361/202449763},
archivePrefix = {arXiv},
       eprint = {2404.10486},
 primaryClass = {astro-ph.GA},
       adsurl = {https://ui.adsabs.harvard.edu/abs/2024A&A...686L...2G}
}

@ARTICLE{el-badry_bh2,
       author = {{El-Badry}, Kareem and {Rix}, Hans-Walter and {Cendes}, Yvette and {Rodriguez}, Antonio C. and {Conroy}, Charlie and {Quataert}, Eliot and {Hawkins}, Keith and {Zari}, Eleonora and {Hobson}, Melissa and {Breivik}, Katelyn and {Rau}, Arne and {Berger}, Edo and {Shahaf}, Sahar and {Seeburger}, Rhys and {Burdge}, Kevin B. and {Latham}, David W. and {Buchhave}, Lars A. and {Bieryla}, Allyson and {Bashi}, Dolev and {Mazeh}, Tsevi and {Faigler}, Simchon},
        title = "{A red giant orbiting a black hole}",
      journal = {\mnras},
         year = 2023,
        month = may,
       volume = {521},
       number = {3},
        pages = {4323-4348},
          doi = {10.1093/mnras/stad799},
archivePrefix = {arXiv},
       eprint = {2302.07880},
 primaryClass = {astro-ph.SR},
       adsurl = {https://ui.adsabs.harvard.edu/abs/2023MNRAS.521.4323E}
}

@ARTICLE{RandoForastier2024,
       author = {{Rando Forastier}, Bruno and {Mar{\'\i}n Pina}, Daniel and {Gieles}, Mark and {Portegies Zwart}, Simon and {Antonini}, Fabio},
        title = "{Binary-single interactions with different mass ratios}",
      journal = {arXiv e-prints},
         year = 2024,
        month = may,
          eid = {arXiv:2405.16999},
        pages = {arXiv:2405.16999},
          doi = {10.48550/arXiv.2405.16999},
archivePrefix = {arXiv},
       eprint = {2405.16999},
 primaryClass = {astro-ph.GA},
       adsurl = {https://ui.adsabs.harvard.edu/abs/2024arXiv240516999R}
}

@ARTICLE{Atallah2024,
       author = {{Atallah}, Dany and {Weatherford}, Newlin C. and {Trani}, Alessandro A. and {Rasio}, Frederic},
        title = "{On Binary Formation from Three Initially Unbound Bodies}",
      journal = {},
         year = 2024,
        month = feb,
          eid = {arXiv:2402.12429},
        pages = {arXiv:2402.12429},
          doi = {10.48550/arXiv.2402.12429},
archivePrefix = {arXiv},
       eprint = {2402.12429},
 primaryClass = {astro-ph.SR},
       adsurl = {https://ui.adsabs.harvard.edu/abs/2024arXiv240212429A}
}

@ARTICLE{galpy,
       author = {{Bovy}, Jo},
        title = "{galpy: A python Library for Galactic Dynamics}",
      journal = {\apjs},
         year = 2015,
        month = feb,
       volume = {216},
       number = {2},
          eid = {29},
        pages = {29},
          doi = {10.1088/0067-0049/216/2/29},
archivePrefix = {arXiv},
       eprint = {1412.3451},
 primaryClass = {astro-ph.GA},
       adsurl = {https://ui.adsabs.harvard.edu/abs/2015ApJS..216...29B}
}

@ARTICLE{Fryer2012,
       author = {{Fryer}, Chris L. and {Belczynski}, Krzysztof and {Wiktorowicz}, Grzegorz and {Dominik}, Michal and {Kalogera}, Vicky and {Holz}, Daniel E.},
        title = "{Compact Remnant Mass Function: Dependence on the Explosion Mechanism and Metallicity}",
      journal = {\apj},
         year = 2012,
        month = apr,
       volume = {749},
       number = {1},
          eid = {91},
        pages = {91},
          doi = {10.1088/0004-637X/749/1/91},
archivePrefix = {arXiv},
       eprint = {1110.1726},
 primaryClass = {astro-ph.SR},
       adsurl = {https://ui.adsabs.harvard.edu/abs/2012ApJ...749...91F}
}

@ARTICLE{MarinPina2024,
       author = {{Mar{\'\i}n Pina}, Daniel and {Rastello}, Sara and {Gieles}, Mark and {Kremer}, Kyle and {Fitzgerald}, Laura and {Rando Forastier}, Bruno},
        title = "{Dynamical formation of Gaia BH3 in the progenitor globular cluster of the ED-2 stream}",
      journal = {\aap},
         year = 2024,
        month = aug,
       volume = {688},
          eid = {L2},
        pages = {L2},
          doi = {10.1051/0004-6361/202450460},
archivePrefix = {arXiv},
       eprint = {2404.13036},
 primaryClass = {astro-ph.GA},
       adsurl = {https://ui.adsabs.harvard.edu/abs/2024A&A...688L...2M}
}

@ARTICLE{Balbinot2024,
       author = {{Balbinot}, E. and {Dodd}, E. and {Matsuno}, T. and {Lardo}, C. and {Helmi}, A. and {Panuzzo}, P. and {Mazeh}, T. and {Holl}, B. and {Caffau}, E. and {Jorissen}, A. and {Babusiaux}, C. and {Gavras}, P. and {Wyrzykowski}, {\L}. and {Eyer}, L. and {Leclerc}, N. and {Bombrun}, A. and {Mowlavi}, N. and {Seabroke}, G.~M. and {Cabrera-Ziri}, I. and {Callingham}, T.~M. and {Ruiz-Lara}, T. and {Starkenburg}, E.},
        title = "{The 33 M$_{{\ensuremath{\odot}}}$ black hole Gaia BH3 is part of the disrupted ED-2 star cluster}",
      journal = {\aap},
         year = 2024,
        month = jul,
       volume = {687},
          eid = {L3},
        pages = {L3},
          doi = {10.1051/0004-6361/202450425},
archivePrefix = {arXiv},
       eprint = {2404.11604},
 primaryClass = {astro-ph.GA},
       adsurl = {https://ui.adsabs.harvard.edu/abs/2024A&A...687L...3B}
}

@ARTICLE{Torres2010,
       author = {{Torres}, Guillermo},
        title = "{On the Use of Empirical Bolometric Corrections for Stars}",
      journal = {\aj},
         year = 2010,
        month = nov,
       volume = {140},
       number = {5},
        pages = {1158-1162},
          doi = {10.1088/0004-6256/140/5/1158},
archivePrefix = {arXiv},
       eprint = {1008.3913},
 primaryClass = {astro-ph.SR},
       adsurl = {https://ui.adsabs.harvard.edu/abs/2010AJ....140.1158T}
}

@ARTICLE{Flower1996,
       author = {{Flower}, Phillip J.},
        title = "{Transformations from Theoretical Hertzsprung-Russell Diagrams to Color-Magnitude Diagrams: Effective Temperatures, B-V Colors, and Bolometric Corrections}",
      journal = {\apj},
         year = 1996,
        month = sep,
       volume = {469},
        pages = {355},
          doi = {10.1086/177785},
       adsurl = {https://ui.adsabs.harvard.edu/abs/1996ApJ...469..355F}
}

@ARTICLE{Shenar2022,
       author = {{Shenar}, Tomer and {Sana}, Hugues and {Mahy}, Laurent and {El-Badry}, Kareem and {Marchant}, Pablo and {Langer}, Norbert and {Hawcroft}, Calum and {Fabry}, Matthias and {Sen}, Koushik and {Almeida}, Leonardo A. and {Abdul-Masih}, Michael and {Bodensteiner}, Julia and {Crowther}, Paul A. and {Gieles}, Mark and {Gromadzki}, Mariusz and {H{\'e}nault-Brunet}, Vincent and {Herrero}, Artemio and {de Koter}, Alex and {Iwanek}, Patryk and {Koz{\l}owski}, Szymon and {Lennon}, Daniel J. and {Ma{\'\i}z Apell{\'a}niz}, Jes{\'u}s and {Mr{\'o}z}, Przemys{\l}aw and {Moffat}, Anthony F.~J. and {Picco}, Annachiara and {Pietrukowicz}, Pawe{\l} and {Poleski}, Rados{\l}aw and {Rybicki}, Krzysztof and {Schneider}, Fabian R.~N. and {Skowron}, Dorota M. and {Skowron}, Jan and {Soszy{\'n}ski}, Igor and {Szyma{\'n}ski}, Micha{\l} K. and {Toonen}, Silvia and {Udalski}, Andrzej and {Ulaczyk}, Krzysztof and {Vink}, Jorick S. and {Wrona}, Marcin},
        title = "{An X-ray-quiet black hole born with a negligible kick in a massive binary within the Large Magellanic Cloud}",
      journal = {Nature Astronomy},
         year = 2022,
        month = jul,
       volume = {6},
        pages = {1085-1092},
          doi = {10.1038/s41550-022-01730-y},
archivePrefix = {arXiv},
       eprint = {2207.07675},
 primaryClass = {astro-ph.HE},
       adsurl = {https://ui.adsabs.harvard.edu/abs/2022NatAs...6.1085S}
}

@ARTICLE{Giesers2018,
       author = {{Giesers}, Benjamin and {Dreizler}, Stefan and {Husser}, Tim-Oliver and {Kamann}, Sebastian and {Anglada Escud{\'e}}, Guillem and {Brinchmann}, Jarle and {Carollo}, C. Marcella and {Roth}, Martin M. and {Weilbacher}, Peter M. and {Wisotzki}, Lutz},
        title = "{A detached stellar-mass black hole candidate in the globular cluster NGC 3201}",
      journal = {\mnras},
         year = 2018,
        month = mar,
       volume = {475},
       number = {1},
        pages = {L15-L19},
          doi = {10.1093/mnrasl/slx203},
archivePrefix = {arXiv},
       eprint = {1801.05642},
 primaryClass = {astro-ph.SR},
       adsurl = {https://ui.adsabs.harvard.edu/abs/2018MNRAS.475L..15G}
}

@ARTICLE{Giesers2019,
       author = {{Giesers}, Benjamin and {Kamann}, Sebastian and {Dreizler}, Stefan and {Husser}, Tim-Oliver and {Askar}, Abbas and {G{\"o}ttgens}, Fabian and {Brinchmann}, Jarle and {Latour}, Marilyn and {Weilbacher}, Peter M. and {Wendt}, Martin and {Roth}, Martin M.},
        title = "{A stellar census in globular clusters with MUSE: Binaries in NGC 3201}",
      journal = {\aap},
         year = 2019,
        month = dec,
       volume = {632},
          eid = {A3},
        pages = {A3},
          doi = {10.1051/0004-6361/201936203},
archivePrefix = {arXiv},
       eprint = {1909.04050},
 primaryClass = {astro-ph.SR},
       adsurl = {https://ui.adsabs.harvard.edu/abs/2019A&A...632A...3G}
}

@ARTICLE{Saracino2022,
       author = {{Saracino}, S. and {Kamann}, S. and {Guarcello}, M.~G. and {Usher}, C. and {Bastian}, N. and {Cabrera-Ziri}, I. and {Gieles}, M. and {Dreizler}, S. and {Da Costa}, G.~S. and {Husser}, T.-O. and {H{\'e}nault-Brunet}, V.},
        title = "{A black hole detected in the young massive LMC cluster NGC 1850}",
      journal = {\mnras},
         year = 2022,
        month = apr,
       volume = {511},
       number = {2},
        pages = {2914-2924},
          doi = {10.1093/mnras/stab3159},
archivePrefix = {arXiv},
       eprint = {2111.06506},
 primaryClass = {astro-ph.GA},
       adsurl = {https://ui.adsabs.harvard.edu/abs/2022MNRAS.511.2914S}
}

@ARTICLE{Mahy2022,
       author = {{Mahy}, L. and {Sana}, H. and {Shenar}, T. and {Sen}, K. and {Langer}, N. and {Marchant}, P. and {Abdul-Masih}, M. and {Banyard}, G. and {Bodensteiner}, J. and {Bowman}, D.~M. and {Dsilva}, K. and {Fabry}, M. and {Hawcroft}, C. and {Janssens}, S. and {Van Reeth}, T. and {Eldridge}, C.},
        title = "{Identifying quiescent compact objects in massive Galactic single-lined spectroscopic binaries}",
      journal = {\aap},
         year = 2022,
        month = aug,
       volume = {664},
          eid = {A159},
        pages = {A159},
          doi = {10.1051/0004-6361/202243147},
archivePrefix = {arXiv},
       eprint = {2207.07752},
 primaryClass = {astro-ph.SR},
       adsurl = {https://ui.adsabs.harvard.edu/abs/2022A&A...664A.159M}
}

@ARTICLE{Sana2012,
       author = {{Sana}, H. and {de Mink}, S.~E. and {de Koter}, A. and {Langer}, N. and {Evans}, C.~J. and {Gieles}, M. and {Gosset}, E. and {Izzard}, R.~G. and {Le Bouquin}, J.-B. and {Schneider}, F.~R.~N.},
        title = "{Binary Interaction Dominates the Evolution of Massive Stars}",
      journal = {Science},
         year = 2012,
        month = jul,
       volume = {337},
       number = {6093},
        pages = {444},
          doi = {10.1126/science.1223344},
archivePrefix = {arXiv},
       eprint = {1207.6397},
 primaryClass = {astro-ph.SR},
       adsurl = {https://ui.adsabs.harvard.edu/abs/2012Sci...337..444S}
}

@ARTICLE{Bressan2012,
       author = {{Bressan}, Alessandro and {Marigo}, Paola and {Girardi}, L{\'e}o. and {Salasnich}, Bernardo and {Dal Cero}, Claudia and {Rubele}, Stefano and {Nanni}, Ambra},
        title = "{PARSEC: stellar tracks and isochrones with the PAdova and TRieste Stellar Evolution Code}",
      journal = {\mnras},
         year = 2012,
        month = nov,
       volume = {427},
       number = {1},
        pages = {127-145},
          doi = {10.1111/j.1365-2966.2012.21948.x},
archivePrefix = {arXiv},
       eprint = {1208.4498},
 primaryClass = {astro-ph.SR},
       adsurl = {https://ui.adsabs.harvard.edu/abs/2012MNRAS.427..127B}
}

@ARTICLE{Costa2025,
       author = {{Costa}, G. and {Shepherd}, K.~G. and {Bressan}, A. and {Addari}, F. and {Chen}, Y. and {Fu}, X. and {Volpato}, G. and {Nguyen}, C.~T. and {Girardi}, L. and {Marigo}, P. and {Mazzi}, A. and {Pastorelli}, G. and {Trabucchi}, M. and {Bossini}, D. and {Zaggia}, S.},
        title = "{Evolutionary tracks, ejecta, and ionizing photons from intermediate-mass to very massive stars with PARSEC}",
      journal = {\aap},
         year = 2025,
        month = feb,
       volume = {694},
          eid = {A193},
        pages = {A193},
          doi = {10.1051/0004-6361/202452573},
archivePrefix = {arXiv},
       eprint = {2501.12917},
 primaryClass = {astro-ph.SR},
       adsurl = {https://ui.adsabs.harvard.edu/abs/2025A&A...694A.193C}
}

@ARTICLE{Nguyen2025,
       author = {{Nguyen}, C.~T. and {Costa}, G. and {Bressan}, A. and {Girardi}, L. and {Cescutti}, G. and {Korn}, A.~J. and {Volpato}, G. and {Chen}, Y. and {Pastorelli}, G. and {Trabucchi}, M. and {Shepherd}, K.~G. and {Ettorre}, G. and {Zaggia}, S.},
        title = "{PARSEC V2.0: Rotating tracks and isochrones for seven additional metallicities in the range Z = 0.0001─0.03}",
      journal = {\aap},
         year = 2025,
        month = sep,
       volume = {701},
          eid = {A258},
        pages = {A258},
          doi = {10.1051/0004-6361/202556005},
archivePrefix = {arXiv},
       eprint = {2508.02393},
 primaryClass = {astro-ph.SR},
       adsurl = {https://ui.adsabs.harvard.edu/abs/2025A&A...701A.258N}
}

@ARTICLE{SchiebelbeinZwack2026,
       author = {{Schiebelbein-Zwack}, Aryanna and {Ye}, Claire S. and {Reina-Campos}, Marta and {Munawwarah}, Aeysha and {El-Badry}, Kareem and {Nagarajan}, Pranav},
        title = "{The Contribution of Disrupted Dense Star Clusters to Gaia's Compact Object Binaries}",
      journal = {arXiv e-prints},
         year = 2026,
        month = jun,
          eid = {arXiv:2606.18419},
        pages = {arXiv:2606.18419},
archivePrefix = {arXiv},
       eprint = {2606.18419},
 primaryClass = {astro-ph.GA},
       adsurl = {https://ui.adsabs.harvard.edu/abs/2026arXiv260618419S}
}

@ARTICLE{MullerHorn2026,
       author = {{M{\"u}ller-Horn}, Johanna and {Rix}, Hans-Walter and {El-Badry}, Kareem and {Pennell}, Ben and {Green}, Matthew and {Li}, Jiadong and {Seeburger}, Rhys},
        title = "{Dormant black hole candidates from Gaia DR3 summary diagnostics}",
      journal = {\aap},
         year = 2026,
        month = may,
       volume = {709},
          eid = {A62},
        pages = {A62},
          doi = {10.1051/0004-6361/202557587},
archivePrefix = {arXiv},
       eprint = {2510.05982},
 primaryClass = {astro-ph.SR},
       adsurl = {https://ui.adsabs.harvard.edu/abs/2026A&A...709A..62M}
}

@ARTICLE{Heggie1996,
       author = {{Heggie}, Douglas C. and {Rasio}, Frederic A.},
        title = "{The Effect of Encounters on the Eccentricity of Binaries in Clusters}",
      journal = {\mnras},
         year = 1996,
        month = oct,
       volume = {282},
       number = {3},
        pages = {1064-1084},
          doi = {10.1093/mnras/282.3.1064},
archivePrefix = {arXiv},
       eprint = {astro-ph/9506082},
 primaryClass = {astro-ph},
       adsurl = {https://ui.adsabs.harvard.edu/abs/1996MNRAS.282.1064H}
}

@ARTICLE{FronimosPouliasis2026,
       author = {{Fronimos Pouliasis}, Fotios and {Dickson}, Nolan and {Mar{\'\i}n Pina}, Daniel and {Gieles}, Mark and {H{\'e}nault-Brunet}, Vincent and {Antonini}, Fabio},
        title = "{cBHBd: A fast code for the evolution of tidally limited star clusters and their binary black hole mergers}",
      journal = {arXiv e-prints},
         year = 2026,
        month = may,
          eid = {arXiv:2605.28088},
        pages = {arXiv:2605.28088},
          doi = {10.48550/arXiv.2605.28088},
archivePrefix = {arXiv},
       eprint = {2605.28088},
 primaryClass = {astro-ph.GA},
       adsurl = {https://ui.adsabs.harvard.edu/abs/2026arXiv260528088F}
}

@ARTICLE{gaiamock,
       author = {{El-Badry}, Kareem and {Lam}, Casey and {Holl}, Berry and {Halbwachs}, Jean-Louis and {Rix}, Hans-Walter and {Mazeh}, Tsevi and {Shahaf}, Sahar},
        title = "{A generative model for Gaia astrometric orbit catalogs: selection functions for binary stars, giant planets, and compact object companions}",
      journal = {The Open Journal of Astrophysics},
         year = 2024,
        month = nov,
       volume = {7},
          eid = {100},
        pages = {100},
          doi = {10.33232/001c.125461},
archivePrefix = {arXiv},
       eprint = {2411.00088},
 primaryClass = {astro-ph.SR},
       adsurl = {https://ui.adsabs.harvard.edu/abs/2024OJAp....7E.100E}
}

@ARTICLE{MarinPina2025a,
       author = {{Mar{\'\i}n Pina}, Daniel and {Gieles}, Mark and {Andrade}, Tomas and {Trani}, Alessandro A.},
        title = "{Interactions among binary black holes in star clusters: Eccentric gravitational wave captures and triple formation}",
      journal = {\aap},
         year = 2025,
        month = jun,
       volume = {698},
          eid = {A229},
        pages = {A229},
          doi = {10.1051/0004-6361/202453531},
archivePrefix = {arXiv},
       eprint = {2501.02907},
 primaryClass = {astro-ph.GA},
       adsurl = {https://ui.adsabs.harvard.edu/abs/2025A&A...698A.229M}
}

@ARTICLE{Jeans1924,
       author = {{Jeans}, J.~H.},
        title = "{Cosmogonic problems associated with a secular decrease of mass}",
      journal = {\mnras},
         year = 1924,
        month = nov,
       volume = {85},
        pages = {2},
          doi = {10.1093/mnras/85.1.2},
       adsurl = {https://ui.adsabs.harvard.edu/abs/1924MNRAS..85....2J}
}

@ARTICLE{Jeans1925,
       author = {{Jeans}, J.~H.},
        title = "{The effect of varying mass on a binary system}",
      journal = {\mnras},
         year = 1925,
        month = jun,
       volume = {85},
        pages = {912},
          doi = {10.1093/mnras/85.9.912},
       adsurl = {https://ui.adsabs.harvard.edu/abs/1925MNRAS..85..912J}
}

@ARTICLE{Mapelli2026,
       author = {{Mapelli}, Michela and {Sgalletta}, Cecilia and {M{\"u}ller-Horn}, Johanna and {Iorio}, Giuliano and {Rinaldi}, Stefano and {Burt}, Christian and {Mar{\'\i}n Pina}, Daniel and {Romagnolo}, Amedeo},
        title = "{The role of accretion efficiency, natal kicks, and angular momentum transport in the formation of the Gaia black holes}",
      journal = {\aap},
         year = 2026,
        month = sep,
       volume = {713},
          eid = {A58},
        pages = {A58},
          doi = {10.1051/0004-6361/202660419},
archivePrefix = {arXiv},
       eprint = {2604.12839},
 primaryClass = {astro-ph.HE},
       adsurl = {https://ui.adsabs.harvard.edu/abs/2026A&A...713A..58M}
}

@ARTICLE{Cavallo2024,
       author = {{Cavallo}, Lorenzo and {Spina}, Lorenzo and {Carraro}, Giovanni and {Magrini}, Laura and {Poggio}, Eloisa and {Cantat-Gaudin}, Tristan and {Pasquato}, Mario and {Lucatello}, Sara and {Ortolani}, Sergio and {Schiappacasse-Ulloa}, Jose},
        title = "{Parameter Estimation for Open Clusters using an Artificial Neural Network with a QuadTree-based Feature Extractor}",
      journal = {\aj},
         year = 2024,
        month = jan,
       volume = {167},
       number = {1},
          eid = {12},
        pages = {12},
          doi = {10.3847/1538-3881/ad07e5},
archivePrefix = {arXiv},
       eprint = {2311.03009},
 primaryClass = {astro-ph.GA},
       adsurl = {https://ui.adsabs.harvard.edu/abs/2024AJ....167...12C}
}

@ARTICLE{Nagarajan2025,
       author = {{Nagarajan}, Pranav and {El-Badry}, Kareem and {Chawla}, Chirag and {Di Carlo}, Ugo Niccol{\`o} and {Breivik}, Katelyn and {Rodriguez}, Carl L. and {Agrawal}, Poojan and {Delfavero}, Vera and {Chatterjee}, Sourav},
        title = "{Realistic Predictions for Gaia Black Hole Discoveries: Comparison of Isolated Binary and Dynamical Formation Models}",
      journal = {\pasp},
         year = 2025,
        month = apr,
       volume = {137},
       number = {4},
          eid = {044202},
        pages = {044202},
          doi = {10.1088/1538-3873/adc839},
archivePrefix = {arXiv},
       eprint = {2502.03527},
 primaryClass = {astro-ph.GA},
       adsurl = {https://ui.adsabs.harvard.edu/abs/2025PASP..137d4202N}
}

@ARTICLE{Anders2021,
       author = {{Anders}, Friedrich and {Cantat-Gaudin}, Tristan and {Quadrino-Lodoso}, Irene and {Gieles}, Mark and {Jordi}, Carme and {Castro-Ginard}, Alfred and {Balaguer-N{\'u}{\~n}ez}, Lola},
        title = "{The star cluster age function in the Galactic disc with Gaia DR2. Fewer old clusters and a low cluster formation efficiency}",
      journal = {\aap},
         year = 2021,
        month = jan,
       volume = {645},
          eid = {L2},
        pages = {L2},
          doi = {10.1051/0004-6361/202038532},
archivePrefix = {arXiv},
       eprint = {2006.01690},
 primaryClass = {astro-ph.GA},
       adsurl = {https://ui.adsabs.harvard.edu/abs/2021A&A...645L...2A}
}

@ARTICLE{Banerjee2010,
       author = {{Banerjee}, Sambaran and {Baumgardt}, Holger and {Kroupa}, Pavel},
        title = "{Stellar-mass black holes in star clusters: implications for gravitational wave radiation}",
      journal = {\mnras},
         year = 2010,
        month = feb,
       volume = {402},
       number = {1},
        pages = {371-380},
          doi = {10.1111/j.1365-2966.2009.15880.x},
archivePrefix = {arXiv},
       eprint = {0910.3954},
 primaryClass = {astro-ph.SR},
       adsurl = {https://ui.adsabs.harvard.edu/abs/2010MNRAS.402..371B}
}

@ARTICLE{Tanikawa2013,
       author = {{Tanikawa}, A.},
        title = "{Dynamical evolution of stellar mass black holes in dense stellar clusters: estimate for merger rate of binary black holes originating from globular clusters}",
      journal = {\mnras},
         year = 2013,
        month = oct,
       volume = {435},
       number = {2},
        pages = {1358-1375},
          doi = {10.1093/mnras/stt1380},
archivePrefix = {arXiv},
       eprint = {1307.6268},
 primaryClass = {astro-ph.GA},
       adsurl = {https://ui.adsabs.harvard.edu/abs/2013MNRAS.435.1358T}
}

@ARTICLE{Ziosi2014,
       author = {{Ziosi}, Brunetto Marco and {Mapelli}, Michela and {Branchesi}, Marica and {Tormen}, Giuseppe},
        title = "{Dynamics of stellar black holes in young star clusters with different metallicities - II. Black hole-black hole binaries}",
      journal = {\mnras},
         year = 2014,
        month = jul,
       volume = {441},
       number = {4},
        pages = {3703-3717},
          doi = {10.1093/mnras/stu824},
archivePrefix = {arXiv},
       eprint = {1404.7147},
 primaryClass = {astro-ph.GA},
       adsurl = {https://ui.adsabs.harvard.edu/abs/2014MNRAS.441.3703Z}
}

@ARTICLE{Mapelli2016,
       author = {{Mapelli}, Michela},
        title = "{Massive black hole binaries from runaway collisions: the impact of metallicity}",
      journal = {\mnras},
         year = 2016,
        month = jul,
       volume = {459},
       number = {4},
        pages = {3432-3446},
          doi = {10.1093/mnras/stw869},
archivePrefix = {arXiv},
       eprint = {1604.03559},
 primaryClass = {astro-ph.GA},
       adsurl = {https://ui.adsabs.harvard.edu/abs/2016MNRAS.459.3432M}
}

@ARTICLE{Chakrabarti2023,
       author = {{Chakrabarti}, Sukanya and {Simon}, Joshua D. and {Craig}, Peter A. and {Reggiani}, Henrique and {Brandt}, Timothy D. and {Guhathakurta}, Puragra and {Dalba}, Paul A. and {Kirby}, Evan N. and {Chang}, Philip and {Hey}, Daniel R. and {Savino}, Alessandro and {Geha}, Marla and {Thompson}, Ian B.},
        title = "{A Noninteracting Galactic Black Hole Candidate in a Binary System with a Main-sequence Star}",
      journal = {\aj},
         year = 2023,
        month = jul,
       volume = {166},
       number = {1},
          eid = {6},
        pages = {6},
          doi = {10.3847/1538-3881/accf21},
archivePrefix = {arXiv},
       eprint = {2210.05003},
 primaryClass = {astro-ph.GA},
       adsurl = {https://ui.adsabs.harvard.edu/abs/2023AJ....166....6C}
}

@ARTICLE{Torniamenti2021,
       author = {{Torniamenti}, Stefano and {Ballone}, Alessandro and {Mapelli}, Michela and {Gaspari}, Nicola and {Di Carlo}, Ugo N. and {Rastello}, Sara and {Giacobbo}, Nicola and {Pasquato}, Mario},
        title = "{The impact of binaries on the evolution of star clusters from turbulent molecular clouds}",
      journal = {\mnras},
         year = 2021,
        month = oct,
       volume = {507},
       number = {2},
        pages = {2253-2266},
          doi = {10.1093/mnras/stab2238},
archivePrefix = {arXiv},
       eprint = {2104.12781},
 primaryClass = {astro-ph.GA},
       adsurl = {https://ui.adsabs.harvard.edu/abs/2021MNRAS.507.2253T}
}

@ARTICLE{Mestichelli2026b,
       author = {{Mestichelli}, Benedetta and {Arca Sedda}, Manuel and {Volonteri}, Marta and {Mapelli}, Michela and {Torniamenti}, Stefano and {Lupi}, Alessandro and {Branchesi}, Marica and {Hirano}, Shingo and {Ishiyama}, Tomoaki and {Klessen}, Ralf S. and {Lipatova}, Veronika and {Liu}, Boyuan},
        title = "{Pebbles to Gems: Intermediate-mass black holes in the first star clusters}",
      journal = {arXiv e-prints},
         year = 2026,
        month = jul,
          eid = {arXiv:2607.03536},
        pages = {arXiv:2607.03536},
          doi = {10.48550/arXiv.2607.03536},
archivePrefix = {arXiv},
       eprint = {2607.03536},
 primaryClass = {astro-ph.GA},
       adsurl = {https://ui.adsabs.harvard.edu/abs/2026arXiv260703536M}
}

@ARTICLE{Kouwenhoven2007,
       author = {{Kouwenhoven}, M.~B.~N. and {Brown}, A.~G.~A. and {Portegies Zwart}, S.~F. and {Kaper}, L.},
        title = "{The primordial binary population. II.. Recovering the binary population for intermediate mass stars in Scorpius OB2}",
      journal = {\aap},
         year = 2007,
        month = oct,
       volume = {474},
       number = {1},
        pages = {77-104},
          doi = {10.1051/0004-6361:20077719},
archivePrefix = {arXiv},
       eprint = {0707.2746},
 primaryClass = {astro-ph},
       adsurl = {https://ui.adsabs.harvard.edu/abs/2007A&A...474...77K}
}

@ARTICLE{Raghavan2010,
       author = {{Raghavan}, Deepak and {McAlister}, Harold A. and {Henry}, Todd J. and {Latham}, David W. and {Marcy}, Geoffrey W. and {Mason}, Brian D. and {Gies}, Douglas R. and {White}, Russel J. and {ten Brummelaar}, Theo A.},
        title = "{A Survey of Stellar Families: Multiplicity of Solar-type Stars}",
      journal = {\apjs},
         year = 2010,
        month = sep,
       volume = {190},
       number = {1},
        pages = {1-42},
          doi = {10.1088/0067-0049/190/1/1},
archivePrefix = {arXiv},
       eprint = {1007.0414},
 primaryClass = {astro-ph.SR},
       adsurl = {https://ui.adsabs.harvard.edu/abs/2010ApJS..190....1R}
}

@ARTICLE{Hackshaw2025,
       author = {{Hackshaw}, Zoe and {Hawkins}, Keith and {Manea}, Catherine},
        title = "{A Detailed Chemical Analysis of the Red Giant Orbiting Gaia BH3: From Lithium to Thorium}",
      journal = {\apj},
         year = 2025,
        month = dec,
       volume = {995},
       number = {2},
          eid = {140},
        pages = {140},
          doi = {10.3847/1538-4357/ae1617},
archivePrefix = {arXiv},
       eprint = {2510.18757},
 primaryClass = {astro-ph.GA},
       adsurl = {https://ui.adsabs.harvard.edu/abs/2025ApJ...995..140H}
}

\onecolumn
\begin{appendix}
\section{Corner plots for additional subsamples}
\begin{figure*}[h]
    \sidecaption
    \includegraphics[width=12cm]{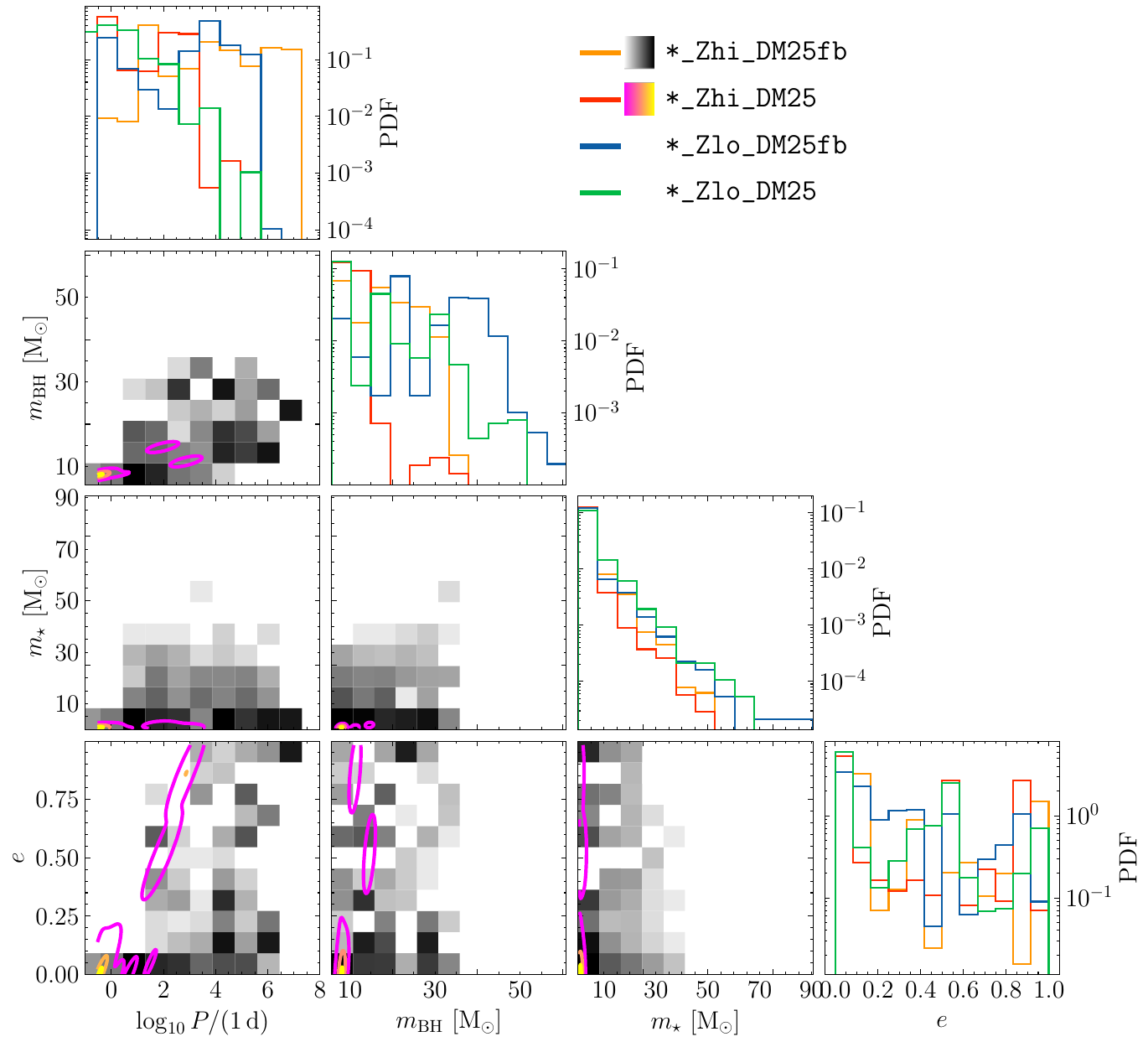}
    \caption{PDF of the orbital period ($P$), BH mass ($m_{BH}$), star mass ($m_\star$), and eccentricity ($e$), for the dormant BHs formed from evolving in isolation the primordial binary population in our cluster models (see Sect.~\ref{sec:population:iso}). The 2D histograms show the high-metallicity models, with DM25fb kicks as a gradient, and DM25 kicks as a contour. The 1D histograms also include the low-metallicity models.}
    \label{fig:corner:iso_all}
\end{figure*}

\begin{figure*}
    \sidecaption
    \includegraphics[width=12cm]{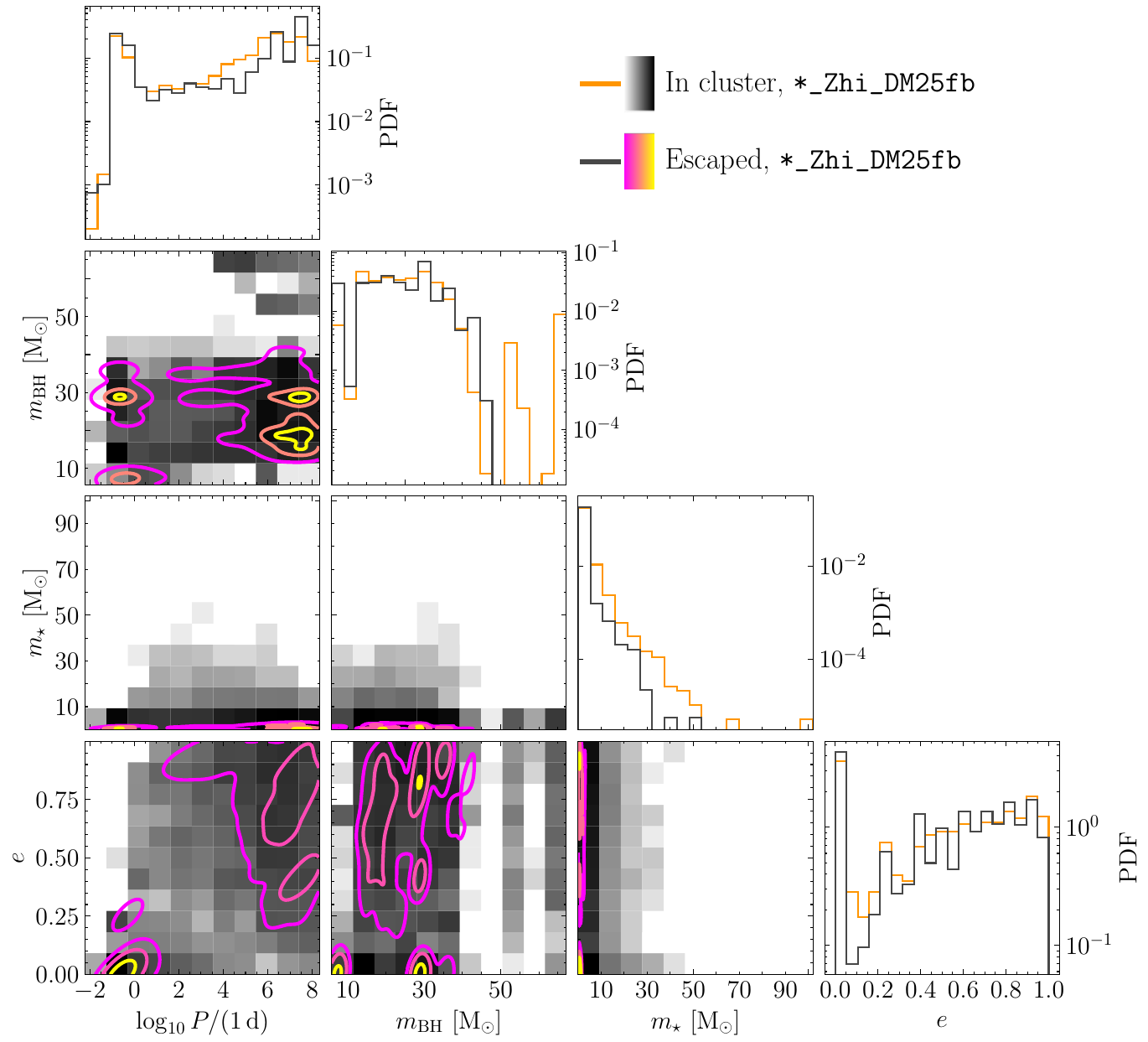}
    \caption{Escaped versus in-cluster dormant BHs (see Sect.~\ref{sec:population:incl}) in model \texttt{_Zhi_DM25fb}. The histograms show the PDF of the orbital period ($P$), BH mass ($m_{BH}$), star mass ($m_\star$), and eccentricity ($e$), for the dormant BHs found in our model with high metallicity and DM25fb kicks.}
    \label{fig:corner:esc_incl}
\end{figure*}

\begin{figure*}
    \sidecaption
    \includegraphics[width=12cm]{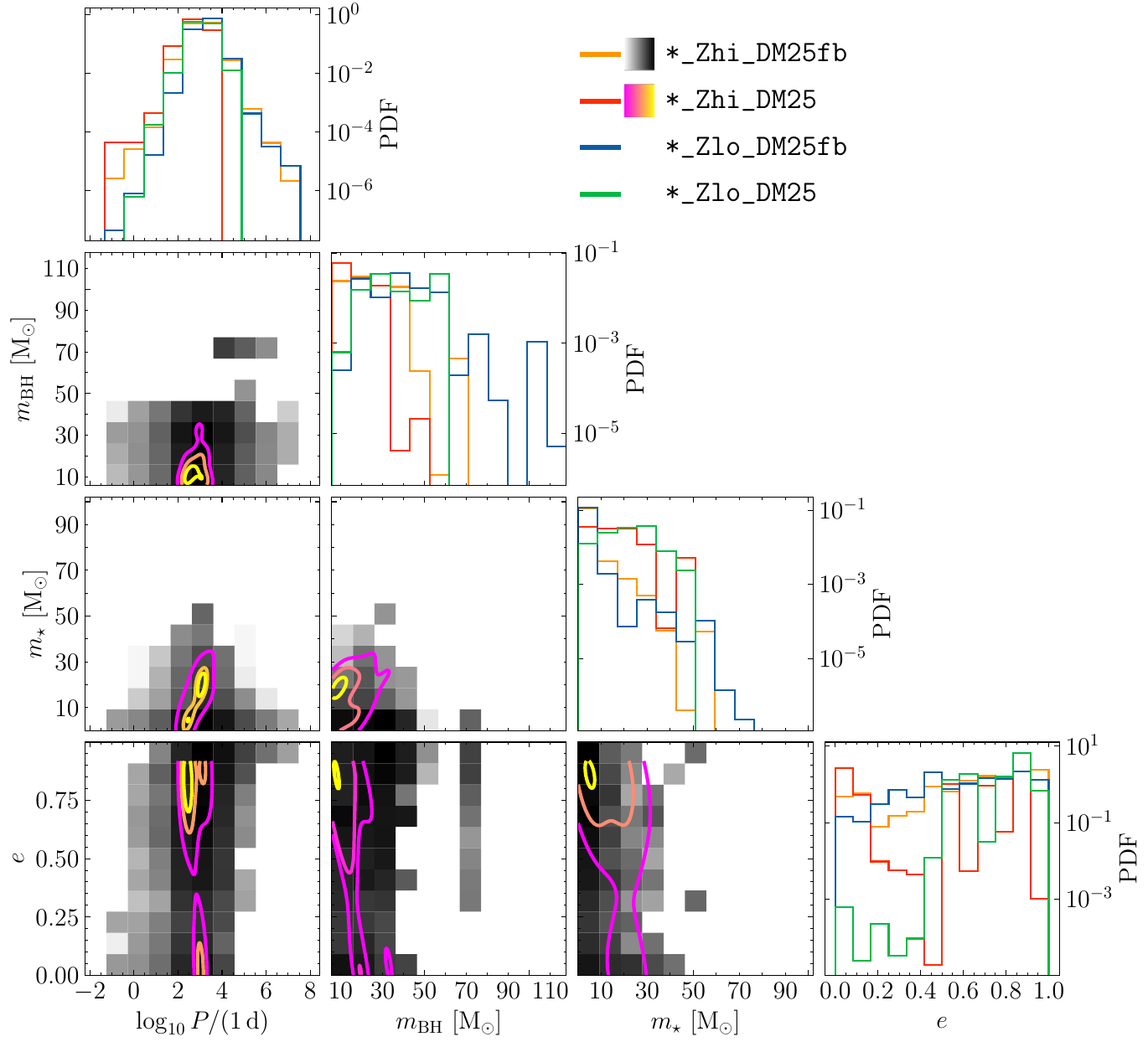}
    \caption{Properties of dormant BHs detectable in \textit{Gaia} DR4. The histograms show the PDF of the orbital period ($P$), BH mass ($m_{BH}$), star mass ($m_\star$), and eccentricity ($e$), for all the dormant BHs found in our simulation, weighted by the detectability in \textit{Gaia} DR4 (see Sect.~\ref{sec:dr4}). The 2D histograms show the high-metallicity models, with DM25fb kicks as a gradient, and DM25 kicks as a contour. The 1D histograms also include the low-metallicity models.}
    \label{fig:corner:vis_all}
\end{figure*}

In this Appendix, we report the corner plots of $\{\log P, m_\BH, m_\star, e\}$ for all the dormant BHs in our simulations, according to different sub-sampling selections.  Figure~\ref{fig:corner:iso_all} shows the dormant BHs formed from evolving in isolation the primordial binary population in our cluster models (see Sect.~\ref{sec:population:iso}). Figure~\ref{fig:corner:esc_incl} shows the dormant BHs in our model with high metallicity and DM25fb kicks, separated in in-cluster and ejected (see Sect.~\ref{sec:population:incl}). Figure~\ref{fig:corner:vis_all} shows all dormant BHs in our simulations weighted by their visibility in DR4 (see Sect.~\ref{sec:dr4}).

\end{appendix}
\end{document}